\documentclass[twocolumn,amsmath,floatfix]{revtex4}
\usepackage{latexsym}
\usepackage{amssymb}
\usepackage{pdfpages}
\usepackage{graphics}

\begin{document}
\newcommand{\ja}{Jakubassa-Amundsen}

\newcommand{\bfx}{\mbox{\boldmath $x$}}
\newcommand{\bfq}{\mbox{\boldmath $q$}}
\newcommand{\bfnabla}{\mbox{\boldmath $\nabla$}}
\newcommand{\bfsigma}{\mbox{\boldmath $\sigma$}}
\newcommand{\bfSigma}{\mbox{\boldmath $\Sigma$}}
\newcommand{\bfsigmas}{\mbox{{\scriptsize \boldmath $\sigma$}}}
\newcommand{\bfeps}{\mbox{\boldmath $\epsilon$}}
\newcommand{\bfGamma}{\mbox{\boldmath $\Gamma$}}
\newcommand{\bfalpha}{\mbox{\boldmath $\alpha$}}
\newcommand{\bfA}{\mbox{\boldmath $A$}}
\newcommand{\bfP}{\mbox{\boldmath $P$}}
\newcommand{\bfF}{\mbox{\boldmath $F$}}
\newcommand{\bfe}{\mbox{\boldmath $e$}}
\newcommand{\bfd}{\mbox{\boldmath $d$}}
\newcommand{\bfes}{\mbox{{\scriptsize \boldmath $e$}}}
\newcommand{\bfn}{\mbox{\boldmath $n$}}
\newcommand{\bfW}{{\mbox{\boldmath $W$}_{\!\!rad}}}
\newcommand{\bfM}{\mbox{\boldmath $M$}}
\newcommand{\bfK}{\mbox{\boldmath $K$}}
\newcommand{\bfI}{\mbox{\boldmath $I$}}
\newcommand{\bfQ}{\mbox{\boldmath $Q$}}
\newcommand{\bfY}{\mbox{\boldmath $Y$}}
\newcommand{\bfp}{\mbox{\boldmath $p$}}
\newcommand{\bfk}{\mbox{\boldmath $k$}}
\newcommand{\bfj}{\mbox{\boldmath $j$}}
\newcommand{\bfks}{\mbox{{\scriptsize \boldmath $k$}}}
\newcommand{\bfs}{\mbox{\boldmath $s$}_0}
\newcommand{\bfv}{\mbox{\boldmath $v$}}
\newcommand{\bfw}{\mbox{\boldmath $w$}}
\newcommand{\bfb}{\mbox{\boldmath $b$}}
\newcommand{\bfxi}{\mbox{\boldmath $\xi$}}
\newcommand{\bfzeta}{\mbox{\boldmath $\zeta$}}
\newcommand{\bfr}{\mbox{\boldmath $r$}}
\newcommand{\bfrs}{\mbox{{\scriptsize \boldmath $r$}}}

\renewcommand{\theequation}{\arabic{section}.\arabic{equation}}
\renewcommand{\thesection}{\arabic{section}}
\renewcommand{\thesubsection}{\arabic{section}.\arabic{subsection}}

\title{\Large\bf Electron--nucleus scattering and the polarization sum rules}

\author{D.~H.~Jakubassa-Amundsen\\
Mathematics Institute, University of Munich, Theresienstrasse 39,\\ 80333 Munich, Germany}



\vspace{1cm}

\begin{abstract}  
A formal derivation of the polarization correlations between the incident electron and the scattered electron is given for a general class of transition operators.
In correspondence to the case of bremsstrahlung emission, three sum rules for the polarization correlations are predicted, which reduce to the known one for potential scattering.
Further examples, including the $1_2^+$ and $2_1^+$ excitations of the $^{12}$C nucleus and a $1^-$ excitation of the $^{208}$Pb nucleus, are discussed and the validity
of the corresponding sum rules is investigated.

\end{abstract}

\maketitle

\vspace{0.5cm}

\section{Introduction}

The differential cross section for the elastic scattering of a polarized electron from a spin-zero nucleus is characterized by three spin asymmetry parameters $R,\;S$ and $L$ (or equivalently, $U,\;S$ and $T$),
which describe the polarization transfer between the incident and the scattered electron. They are 
 accessible by varying  the direction of the polarization vector $\bfzeta_i$ of
the beam electron \cite{Motz}.
Experimentally these polarization correlations can be  obtained 
from measurements at the CEBAF (Jefferson Lab)  or MAMI (Mainz) facilities \cite{Gr20,Sch17,Au18} 
in terms of  relative cross-section differences when the spin of the beam electron is flipped,
\begin{equation}\label{1.1}
P(\bfzeta_i)\;=\;\frac{d\sigma/d\Omega_f(\bfzeta_i)-d\sigma/d\Omega_f(-\bfzeta_i)}{d\sigma/d\Omega_f(\bfzeta_i)+d\sigma/d\Omega_f(-\bfzeta_i)},
\end{equation}
where $P$ stands for one of the parameters $S,\;R$ or $L$.
While for potential scattering these parameters obey the sum rule \cite{Motz}
\begin{equation}\label{1.2}
\Sigma_3\;=\;S^2+R^2+L^2=1,
\end{equation}
  this three-term sum rule is no longer strictly valid for arbitrary scattering operators, in particular for scattering from nuclei carrying spin or for nuclear excitation, if more that one single final nuclear state contributes  \cite{Jaku12,Jaku15}. 

In the case of polarized bremsstrahlung emission, there exist similar polarization correlations \cite{TP73}.
For the parameters entering into the doubly differential bremsstrahlung cross section, a seven-term sum rule was established
under the assumption that the unobserved final electron can be described by just two partial-wave states \cite{PMS}.
Profiting from this result, but allowing for the additional observation of the scattered electron, it was shown that there exist three seven-term
sum rules, provided that the momenta of electron and photon are coplanar \cite{Jaku18}.
Such sum rules can be used to determine spin asymmetries for which an experimental observation is not possible,
but they provide also a stringent test for the theoretical models.

In the present work the parametrization of the cross section for potential scattering \cite{Motz,Jaku12a} is extended to a more general class of electron-nucleus scattering
cross sections by introducing a total set of fifteen polarization correlations.
The results for bremsstrahlung are then used to derive three seven-term sum rules for these polarization correlations.
For the special case of elastic scattering from spin-zero nuclei it is shown that these sum rules reduce to  the  sum rule  (\ref{1.2}).
For inelastic scattering it will be demonstrated that one or two sum rules hold if the excited nucleus can be represented by a superposition of at most two final states.

The paper is organized as follows. Section 2 provides the formal derivation of the polarization correlations for spin-polarized electron impact. The three seven-term sum rules are established in section 3
and are adapted to the  special cases of elastic scattering from spin-zero nuclei and nuclear excitation.
Section 4 furnishes results for the differential cross section in comparison with experiment as well as for the sum rules in the case of  selected dipole
and quadrupole excitations of $^{12}$C and $^{208}$Pb.
Investigations of the numerical accuracy of the sum rules are provided in section 5, including a quadrupole excitation of the medium-heavy $^{92}$Zr nucleus.
Concluding remarks follow (section 6). Atomic units ($\hbar=m=e=1$) are used unless indicated otherwise.

\section{Scattering theory and polarization correlations}
\setcounter{equation}{0}

For the derivation of the scattering formalism we restrict ourselves to the  case where the nucleus is in a fixed
initial and final state without any coupling to the electronic polarization degrees of freedom.
 The differential cross section for the transition of an electron with spin polarization $\bfzeta_i$ into a state with spin polarization $\bfzeta_f$, while being scattered into the solid angle $d\Omega_f$, is given by
$$\frac{d\sigma}{d\Omega_f}(\bfzeta_i,\bfzeta_f)\;=\;N_0\;|W_{fi}|^2,$$
\begin{equation}\label{2.1}
W_{fi}\,=\,\langle \psi_f(\bfzeta_f,\bfr)\,|\hat{O}(\bfr)|\,\psi_i(\bfzeta_i,\bfr)\rangle,
\end{equation}
where $N_0$ is a normalization constant, $\hat{O}$ is some transition operator, and $\psi_i $ and $\psi_f$ describe the initial and final electronic states, respectively, which are linear in the polarization spinors \cite{Rose},
$$\psi_i(\bfzeta_i,\bfr)\,=\,\sum_{m_i=\pm 1/2} a_{m_i}\;\phi_i^{(m_i)}(\bfr)$$
\begin{equation}\label{2.2}
\psi_f^+(\bfzeta_f,\bfr)\,=\,\sum_{m_s=\pm 1/2} b_{m_s}^\ast\;\phi_f^{+(m_s)}(\bfr).
\end{equation}
The coefficients $a_{m_i}$ are related to the spherical coordinates $(1,\alpha_s,\varphi_s)$ of $\bfzeta_i$ by means of
\begin{equation}\label{2.3}
a_{1/2}\,=\,\cos \frac{\alpha_s}{2}\;e^{-i\varphi_s/2},\quad a_{-1/2}\,=\,\sin \frac{\alpha_s}{2}\;e^{i\varphi_s/2}.
\end{equation}
Corresponding relations hold for $b_{m_s}$.
Hence the transition amplitude $W_{fi}$ can be written in the following form,
\begin{equation}\label{2.4}
W_{fi}\,=\,\sum_{m_i=\pm 1/2} a_{m_i}\sum_{m_s=\pm 1/2} b_{m_s}^\ast\;M_{fi}(m_i,m_s).
\end{equation}

Using the short-hand notation,
$$M_{fi}(\frac12,\frac12)=J,\qquad M_{fi}(-\frac12,\frac12)=G,$$
\begin{equation}\label{2.5}
M_{fi}(\frac12,-\frac12)=K,\qquad M_{fi}(-\frac12,-\frac12)=H,
\end{equation}

the cross section turns into
$$\frac{d\sigma}{d\Omega_f}(\bfzeta_i,\bfzeta_f)=N_0\left| a_\frac12 b^\ast_\frac12\, J+a_\frac12 b^\ast_{-\frac12}\,K+a_{-\frac12} b^\ast_\frac12 \,G \right.$$
\begin{equation}\label{2.6}
\left. +a_{-\frac12}b^\ast_{-\frac12}\,H\right|^2.
\end{equation}

By (\ref{2.3}), the coefficients $a_{m_i}$ are also  related to the Cartesian coordinates $(\zeta_{ix},\zeta_{iy},\zeta_{iz}) =(\zeta_{i1},\zeta_{i2},\zeta_{i3})$ of $\bfzeta_i$, in particular one has 
$$|a_{\pm \frac12}|^2\;=\;\frac12 \,(1\pm \cos \alpha_s)\;=\;\frac12\,(1\pm \zeta_{iz})$$
\begin{equation}\label{2.7}
a^\ast_\frac12 a_{-\frac12}\;=\;\frac12\,\sin \alpha_s\,e^{i\varphi_s}\;=\;\frac12\,(\zeta_{ix} + i \zeta_{iy})
\end{equation}
and similarly,
\begin{equation}\label{2.8}
|b_{\pm \frac12}|^2\;=\,\frac12\, (1\pm\zeta_{fz}),\quad \;b^\ast_\frac12 b_{-\frac12}\;=\;\frac12\,(\zeta_{fx}+i\zeta_{fy}).
\end{equation}
Thus the differential cross section is at most bilinear in the coordinates of $\bfzeta_i$ and $\bfzeta_f$,
$$\frac{d\sigma}{d\Omega_f}(\bfzeta_i,\bfzeta_f)\;=\;\frac12\left( \frac{d\sigma}{d\Omega_f} \right)_0 \left[ 1\,+\,\sum_{j=1}^3 c_{j0}\;\zeta_{ij} \right.$$
\begin{equation}\label{2.9}
\left. +\;\sum_{k=1}^3 c_{0k}\;\zeta_{fk}
 +\,\sum_{j,k=1}^3 c_{jk}\;\zeta_{ij}\,\zeta_{fk}\right],
\end{equation}
where $(\frac{d\sigma}{d\Omega_f})_0$ is the differential cross section for unpolarized electrons,
and the parameters $c_{jk}$ define the polarization correlations.
They are obtained by means of evaluating the cross section (\ref{2.6}) and comparing it to its  formal representation (\ref{2.9}).
The result for the 15 $c_{jk}$  is
\begin{eqnarray*}
c_{01}&=&[2\mbox{ Re }(JK^\ast)\,+ 2\mbox{ Re }(GH^\ast)]/D_0\\
c_{02}&=& [-2 \mbox{ Im }(JK^\ast)\,-\,2\mbox{ Im }(GH^\ast)]/D_0\\
c_{03}&=&[\,|J|^2\,-\,|K|^2\,+\,|G|^2\,-\,|H|^2]/D_0\\
&&\\
c_{10}&=& [2\mbox{ Re }(JG^\ast)\,+2\mbox{ Re }(KH^\ast)]/D_0\\
c_{11}&=&[2\mbox{ Re }(JH^\ast)\,+2\mbox{ Re }(GK^\ast)]/D_0\\
c_{12}&=&[-2\mbox{ Im }(JH^\ast)\,-2\mbox{ Im }(GK^\ast)]/D_0\\
c_{13}&=&[2\mbox{ Re }(JG^\ast)\,-2\mbox{ Re }(KH^\ast)]/D_0\\
&&\\
c_{20}&=&[2\mbox{ Im }(JG^\ast)\,+2\mbox{ Im }(KH^\ast)]/D_0\\
c_{21}&=&[2 \mbox{ Im }(JH^\ast)\,-2\mbox{ Im }(GK^\ast)]/D_0\\
c_{22}&=& [2 \mbox{ Re }(JH^\ast) \,-2\mbox{ Re }(GK^\ast)]/D_0\\
c_{23}&=&[2\mbox{ Im }(JG^\ast)\,-2\mbox{ Im }(KH^\ast)]/D_0\\
&&\\
c_{30}&=&[\,|J|^2\,+\,|K|^2\,-\,|G|^2\,-\,|H|^2]/D_0\\
c_{31}&=&[2\mbox{ Re }(JK^\ast)\,-2\mbox{ Re }(GH^\ast)]/D_0\\
c_{32}&=&[-2\mbox{ Im }(JK^\ast)\,+2\mbox{ Im }(GH^\ast)]/D_0
\end{eqnarray*}
\begin{equation}\label{2.10}
c_{33}\,=\,[\,|J|^2\,-\,|K|^2\,-\,|G|^2\,+\,|H|^2]/D_0,
\end{equation}
and for the unpolarized cross section one gets
$$\left( \frac{d\sigma}{d\Omega_f}\right)_0\,=\,\frac12\sum_{\zeta_i,\zeta_f} \frac{d\sigma}{d\Omega_f}(\bfzeta_i,\bfzeta_f)\;=\;\frac12\;N_0\;D_0,$$
\begin{equation}\label{2.11}
D_0\,=\,|J|^2\,+\,|K|^2\,+\,|G|^2\,+\,|H|^2.
\end{equation}
In the absence of time-reversal invariance, all $c_{jk}$ can be  nonzero and distinct.


\section{Sum rules}
\setcounter{equation}{0}

We start by setting up the general sum rules which are obeyed by the parameters defined in (\ref{2.10}).
Subsequently we demonstrate their simplification for elastic scattering from spin-zero nuclei.
Finally we investigate under which conditions some are valid for nuclear excitation.

\subsection{General formulation}

The sum rules given below are based on the results for bremsstrahlung emission in electron-nucleus scattering \cite{PMS,Jaku18}. One of the bremsstrahlung sum rules was obtained for the case of polarized initial and final electrons, but  unpolarized emitted photons in coplanar geometry.
Noting that in the bremsstrahlung polarization correlations, termed  $\tilde{C}_{jkl}$, the indices $j$ and $l$ relate to the components of $\bfzeta_i$ and $\bfzeta_f$, respectively,
while the index $k$ relates to those of the photon, the corresponding sum rule for electron scattering (without photon emission)  is obtained from this bremsstrahlung sum rule by dropping $k$,
i.e. by identifying $c_{jl}=\tilde{C}_{j0l}$ (where $k=0$ denotes the absence of any photon polarization component in the respective contribution to the cross section). This procedure leads to the first seven-term sum rule,
\begin{equation}\label{3.1}
\Sigma_7\;=\;c_{33}^2+(c_{13}^2+c_{20}^2)+(c_{02}^2+c_{31}^2)+c_{11}^2-c_{22}^2=1.
\end{equation}
It can be proved directly from (\ref{2.10}) or equivalently,  by identifying the bremsstrahlung matrix elements $J_+ =M_{fi}(\epsilon^\ast_+,\frac12,\frac12)$ with $J$, $J_-=M_{fi}(\epsilon^\ast_-,\frac12,\frac12)$ with $H$, $S_+=M_{fi}(\epsilon^\ast_+,-\frac12,\frac12)$ with $G$ and $S_-=M_{fi}(\epsilon^\ast_-,-\frac12,\frac12)$ with $-K$,
where $\epsilon_+$ and $\epsilon_-$ denote right-handed and left-handed (circularly polarized) photons, respectively. 
With this identification, the remaining eight $c_{jl}$ can be related to the $\tilde{C}_{jkl}$ with $k=2$ by dropping the second index, $c_{jl}=\tilde{C}_{j2l}$.
Hence, the two other bremsstrahlung sum rules which are associated with  unpolarized beam electrons or unpolarized scattered electrons translate, respectively, to
$$c_{01}^2+c_{02}^2+c_{03}^2+c_{21}^2+c_{22}^2+c_{23}^2-c_{20}^2=1,$$
\begin{equation}\label{3.2}
c_{10}^2+c_{20}^2+c_{30}^2+c_{12}^2+c_{22}^2+c_{32}^2-c_{02}^2=1.
\end{equation}

It is seen that the parameters in the second sum rule of (\ref{3.2}) have reversed indices as compared to those in the first one (i.e. $c_{kj}$ instead of $c_{jk}$). This manifests the symmetry between impinging and outgoing electron and relates to the time-reversal invariance
of the coplanar bremsstrahlung emission.
This fact is also evident in the sum rule (\ref{3.1}) which is symmetric upon index reversion.

For the physical interpretation of the first sum rule in (\ref{3.2}) we take $\bfzeta_i=(0,1,0)$.
With $\zeta_{ix}=\zeta_{iz}=0$, the differential cross section (\ref{2.9}) contains exactly the seven parmeters involved in this sum rule.
The parameters of the second sum rule in (\ref{3.2}) are recovered for the case $\bfzeta_f=(0,1,0).$

\begin{figure}
\vspace{-1.5cm}
\includegraphics[width=11cm]{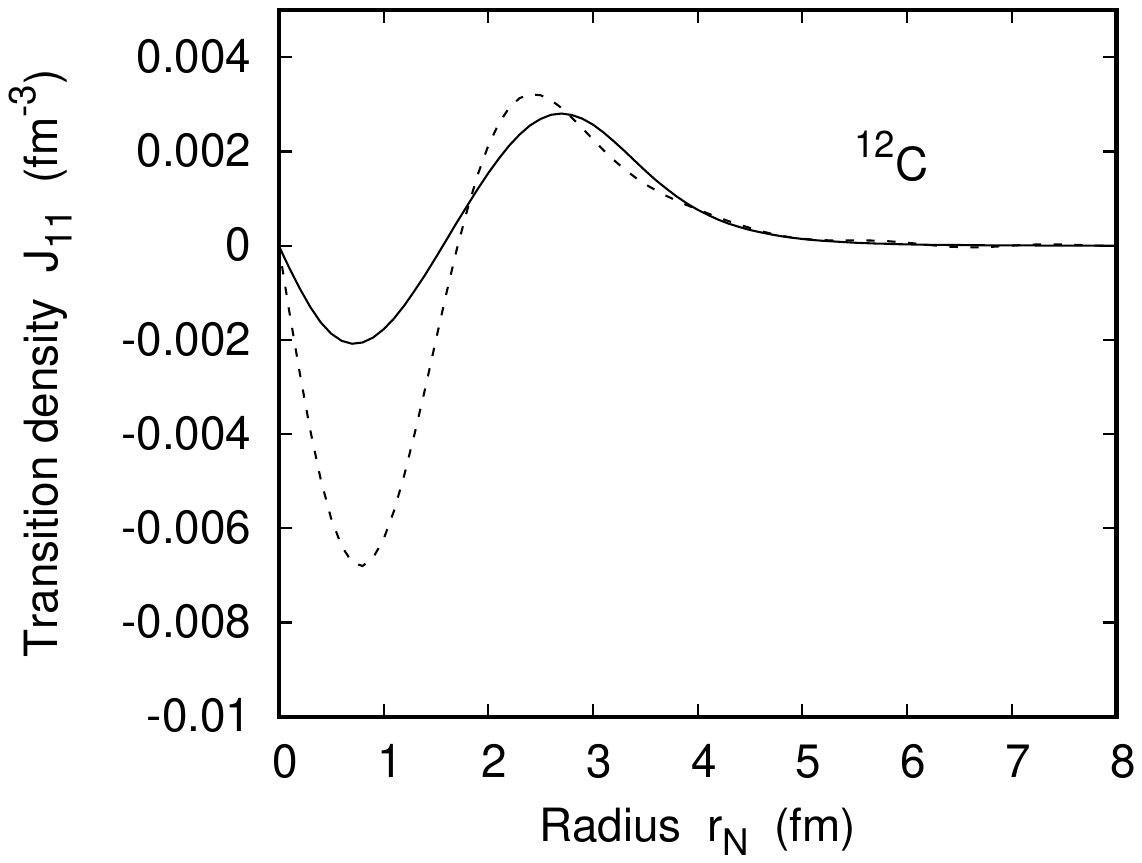}
\vspace{-1.5cm}
\caption
{Current transition density $J_{11}$ for the $1_2^+$ state in $^{12}$C, obtained from the QPM calculation (-----, \cite{Po19}) and from a Fourier-Bessel fit ($-----$, \cite{Deu83}), 
as a function of the distance $r_N$ from the nuclear center.
}
\end{figure}

\subsection{Elastic scattering from spin-zero nuclei}

For Coulombic scattering the interaction between electron and nucleus can be expressed in terms of a potential.
Thus $\hat{O}(r)$ is substituted by the nuclear potential,
while $\phi_i^{(m_i)}(\bfr)$ in (\ref{2.2}) represents an electronic  plane wave and $\phi_f^{(m_s)}(\bfr)$  an exact state.
A coordinate system is chosen with the z-axis along the momentum $\bfk_i$ of the beam electron, and with the momentum $\bfk_f$ of the scattered electron  in the $(x,z)$-plane. The y-axis is taken along $\bfk_i \times \bfk_f$.

Within the phase-shift analysis for potential scattering, the transition amplitude $M_{fi}(m_i,m_s)$
consists of the direct term $A$ and the spin-flip term $B$ \cite{Lan},
\begin{equation}\label{4.1a}
M_{fi}(m_i,m_s)\,=\,A\;\langle \chi_{m_s}|\,\chi_{m_i}\rangle\;+\;B\;\langle \chi_{m_s}|\,\bfn\bfsigma\,|\chi_{m_i}\rangle,
\end{equation}
where $\bfsigma$ is the vector of Pauli spin matrices, $\bfn$ is the normal to the $(x,z)$ scattering plane and $\chi_\frac12={1\choose 0},\;\chi_{-\frac12}={0 \choose 1}$ are the spinor basis states.

From (\ref{4.1a}) the symmetry relations are obtained,
$$M_{fi}(\frac12,\frac12)\,=\,M_{fi}(-\frac12,-\frac12)\,=A$$
\begin{equation}\label{4.2}
M_{fi}(\frac12,-\frac12)\,=\,-M_{fi}(-\frac12,\frac12)\,=\,iB,
\end{equation}
such that $H=J=A$ and $K=-G=iB$.
This implies that the polarization correlations (\ref{2.10}) become mutually dependent,
\begin{equation}\label{4.3}
c_{11}=c_{33},\qquad c_{02}=c_{20},\qquad c_{31}=-c_{13}.
\end{equation}
Explicitly,
$$c_{20}\,=\,\frac{2 \mbox{ Re }(AB^\ast)}{|A|^2+|B|^2},\qquad c_{33}\,=\,\frac{|A|^2-|B|^2}{|A|^2+|B|^2},$$
\begin{equation}\label{4.4}
c_{31}\,=\,\frac{2\mbox{ Im }(AB^\ast)}{|A|^2+|B|^2},\qquad c_{22}\,=\,1,
\end{equation}
while all other $c_{jk}$ are zero. For the unpolarized cross section one gets $(\frac{d\sigma}{d\Omega_f})_0=|A|^2+|B|^2.$ This reduces the sum rule (\ref{3.1}) to a simple three-term sum rule,
\begin{equation}\label{4.5}
c_{31}^2\,+\,c_{20}^2\,+\,c_{33}^2\,=1,
\end{equation}
while the sum rules (\ref{3.2}) become trivial.
With the identification $c_{20}=S,\;c_{33}=T$ and $c_{31}=U$, the result from Motz et al. \cite{Motz} is recovered. 
In fact, (\ref{4.5}) is also easily expressed in terms of $L$ and $R$ by means of $T= L\,\cos\vartheta_f - R\,\sin \vartheta_f$ and $U=L\,\sin \vartheta_f + R \,\cos \vartheta_f$, where $\vartheta_f$ is the scattering angle, and hence coincides with (\ref{1.2}).

\begin{figure}
\vspace{-1.5cm}
\includegraphics[width=11cm]{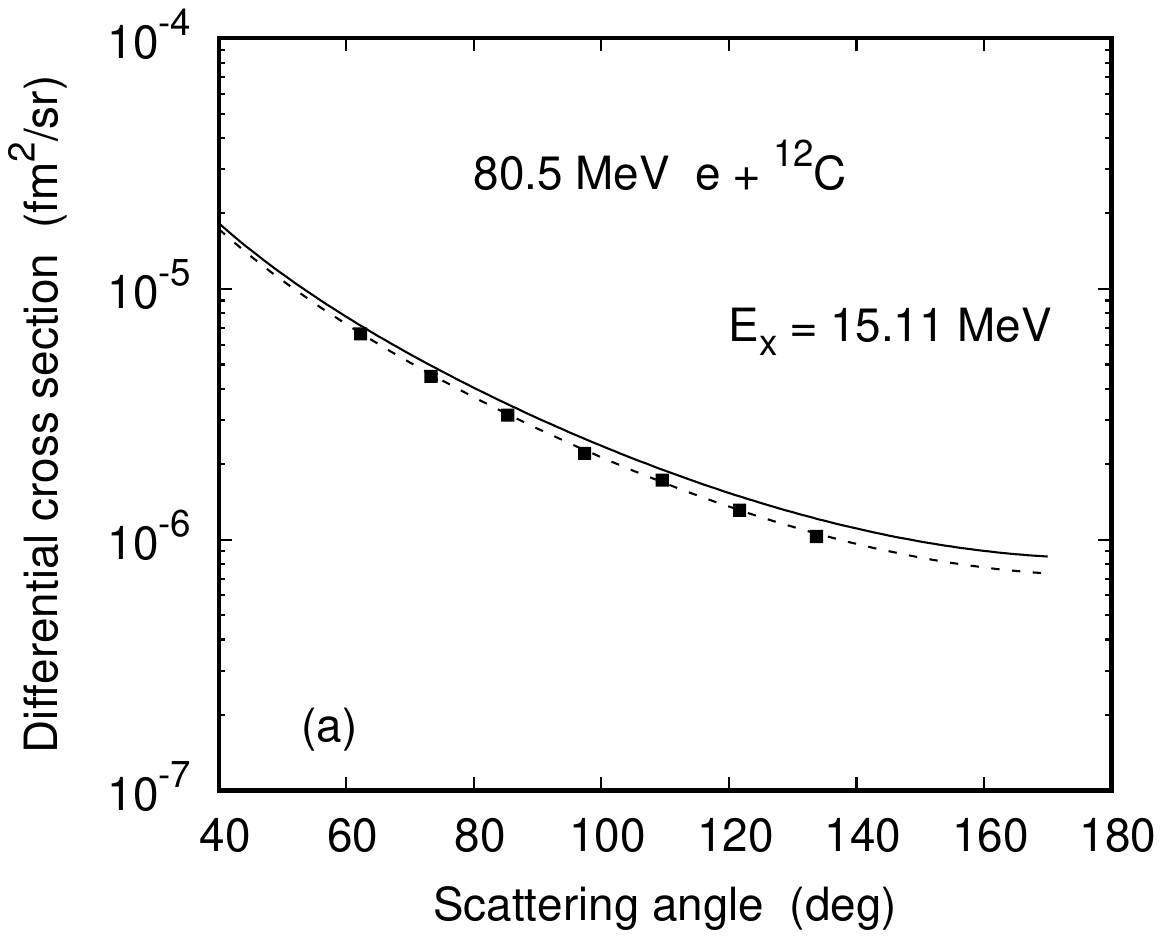}
\vspace{-1.5cm}
\vspace{-0.5cm}
\includegraphics[width=11cm]{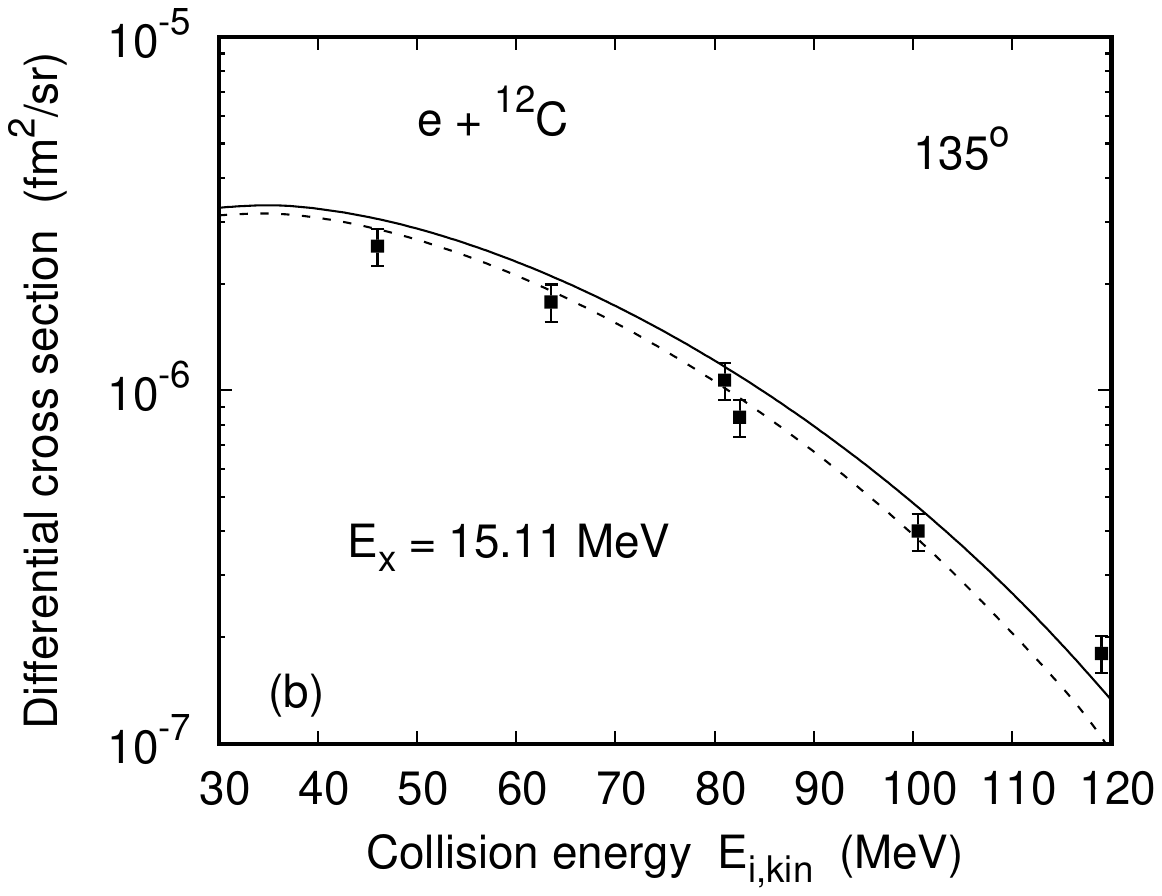}
\caption
{
Differential cross section $\frac{d\sigma}{d\Omega_f}$ for the excitation of the 15.11 MeV state (a)
in 80.5 MeV $e + ^{12}$C collisions as a function of scattering angle $\vartheta_f$
and (b) in $e + ^{12}$C collisions at $\vartheta_f = 135^\circ$ as a function of collision energy $E_{i,kin}=E_i-c^2$.
In (a), the results using the QPM (-----) and the Fourier-Bessel ($-----$) transition density $J_{11}$ are compared
with the experimental data from Deutschmann et al.  ($\blacksquare$, \cite{Deu83}).
In (b), a comparison of the results using the respective densities  with the experimental data from Proca and Isabelle ($\blacksquare$, \cite{PI68}) is given.
}
\end{figure}

\subsection{Nuclear excitation}

The formalism of section 2 can readily be extended to the excitation of a nucleus to a   state with nonzero spin but unobserved spin polarization.
For the sake of demonstration we treat the case of (transverse magnetic) excitation to a  state with unnatural parity.
The respective transition operator is given by \cite{TW68,Jaku15}
\begin{equation}\label{5.1}
\hat{O}(\bfr)\;=\;\frac{\bfalpha}{c} \int d\bfr_N \frac{\stackrel{\leftrightarrow}{I}\,e^{ik|\bfrs-\bfrs_N|}}{|\bfr-\bfr_N|}\;\bfj_{fi}(\bfr_N),
\end{equation}
where $\stackrel{\leftrightarrow}{I}$ is the dyadic unit matrix, 
$\bfalpha$ is a vector of Dirac matrices, $ kc=E_x$ is the nuclear excitation energy, and for transverse magnetic transitions one has
\begin{equation}\label{5.2}
\bfj_{fi}(\bfr_N)\,=\,-i \sum_{LM} (J_iM_iLM|J_fM_f)\;J_{LL}(r_N)\;\bfY_{LL}^{M\ast}(\hat{r}_N).
\end{equation}
The density $J_{LL}(r_N)$ for the transition from the nuclear ground state with angular momentum $J_i$ to an excited state with angular momentum $J_f$, having  parity $\pi_{fi}=(-1)^{L+1}$, is obtained from nuclear models, and
 $\bfY_{LL}^M$ is a vector spherical harmonics \cite{Ed}.

Within the distorted-wave Born approximation (DWBA)
both $\phi_i^{(m_i)}(\bfr)$ and $\phi_f^{(m_s)}(\bfr)$ are scattering eigenstates to the nuclear potential. Their partial-wave expansion is given by
\cite{Rose}
$$\phi_i^{(m_i)}(\bfr)=\sum_{j_il_i} \sqrt{\frac{2l_i+1}{4\pi}}\;(l_i\,0\,\frac12\,m_i|\,j_i\,m_i)
\; i^{l_i}\,e^{i\delta_{\kappa_i}}\;\psi_{\kappa_im_i}(\bfr),$$
$$\phi_f^{+(m_s)}(\bfr)=\sum_{j_fl_fm_f}(l_f\,m_l\,\frac12\,m_s|\,j_f\,m_f)$$
\begin{equation}\label{5.3}
\times\; (-i)^{l_f}\,e^{i\delta_{\kappa_f}}\,Y_{l_fm_l}(\hat{\bfk}_f)\;\psi^+_{\kappa_fm_f}(\bfr),
\end{equation}
with $m_l=m_f-m_s$, the interrelation $j=|\kappa|-\frac12,\;l=|\kappa +\frac12|-\frac12, \;l'=|\kappa -\frac12|-\frac12$ and the phase shifts $\delta_{\kappa_i},\delta_{\kappa_f}$.

The transition matrix element involving the four-spinors $\psi_{\kappa_i m_i},\;\psi_{\kappa_fm_f}$ is
expressed in terms of
 $g_{\kappa_i},g_{\kappa_f}$ and $f_{\kappa_i},f_{\kappa_f}$ which are, respectively,  the large and small components of the radial Dirac eigenfunction,
$$\psi^+_{\kappa_fm_f}(\bfr)\,\bfalpha\,\psi_{\kappa_im_i}(\bfr)\,=\,i\left[ g_{\kappa_f}f_{\kappa_i}(Y^+_{j_fl_fm_f}\bfsigma\,Y_{j_il'_im_i})\right.$$
\begin{equation}\label{5.4}
\left. -\;f_{\kappa_f }g_{\kappa_i}(Y^+_{j_fl'_fm_f}\,\bfsigma\,Y_{j_il_im_i})\right],
\end{equation}
where $Y_{lm}$ is a spherical harmonics and $Y_{jlm}$ is a spherical harmonic spinor \cite{Ed}.
The result for the transition amplitude $M_{fi}(m_i,m_s,M_f)$, where we have included the final nuclear spin polarization $M_f$ as a third variable, is 
in case of the transverse magnetic excitation \cite{Jaku15}
$$M_{fi}^{tm}(m_i,m_s,M_f)=\sum_{l_fm_l} Y_{l_fm_l}(\hat{\bfk}_f) \sum_{j_fm_f} (l_fm_l\,\frac12\,m_s|\,j_fm_f)$$
$$\times \sum_{LM}\sum_{j_il_i} N_{fi}\;(l_i0\,\frac12\,m_i|\,j_im_i)\,(J_iM_iLM|\,J_fM_f)$$
\begin{equation}\label{5.5}
\times\;\left[ R_{12}(g_{\kappa_f}f_{\kappa_i})\,W_0(l_f,l'_i,L)\,-R_{12}(f_{\kappa_f}g_{\kappa_i})\,W_0(l'_f,l_i,L)\right],
\end{equation}
where $R_{12}(g_{\kappa_f}f_{\kappa_i})$ and $R_{12}(f_{\kappa_f}g_{\kappa_i})$ are the radial integrals, and $N_{fi}=i^{l_i-l_f+1} \sqrt{4\pi(2l_i+1)}\; e^{i(\delta_{\kappa_f}+\delta_{\kappa_i})} k/c$.
The angular integral
\begin{equation}\label{5.6}
W_0(l_f,l'_i,L)\,=\,\int d\Omega\;Y^+_{j_fl_fm_f}(\hat{\bfr})\,(\bfY_{LL}^{M\ast}(\hat{\bfr})\bfsigma)\;Y_{j_il'_im_i}(\hat{\bfr})
\end{equation}
is given by (using the conventions of Edmonds \cite{Ed})
$$W_0(l_f,l'_i,L)\,=\,(-1)^M C_0\;(l'_i0L0|\,l_f0) \sum_{m_{s_f}m_{s_i}}$$
$$\times\;\sum_{\mu_f\mu_i}\sum_{\mu \varrho}(l_f\mu_f\,\frac12\,m_{s_f}|\,j_fm_f)\,(l'_i\mu_i\,\frac12\,m_{s_i}|\,j_im_i)$$
\begin{equation}\label{5.7}
\times \;(L\mu 1 \varrho|\,LM)\;(\frac12 \,m_{s_i} 1 -\varrho|\,\frac12\,m_{s_f})\;(l'_i \mu_i L -\mu|\,l_f\mu_f)
\end{equation}
with the constant $C_0=\sqrt{\frac{3(2L+1)(2l'_i+1)}{4\pi (2l_f+1)}}$.

\begin{figure}
\vspace{-1.5cm}
\includegraphics[width=11cm]{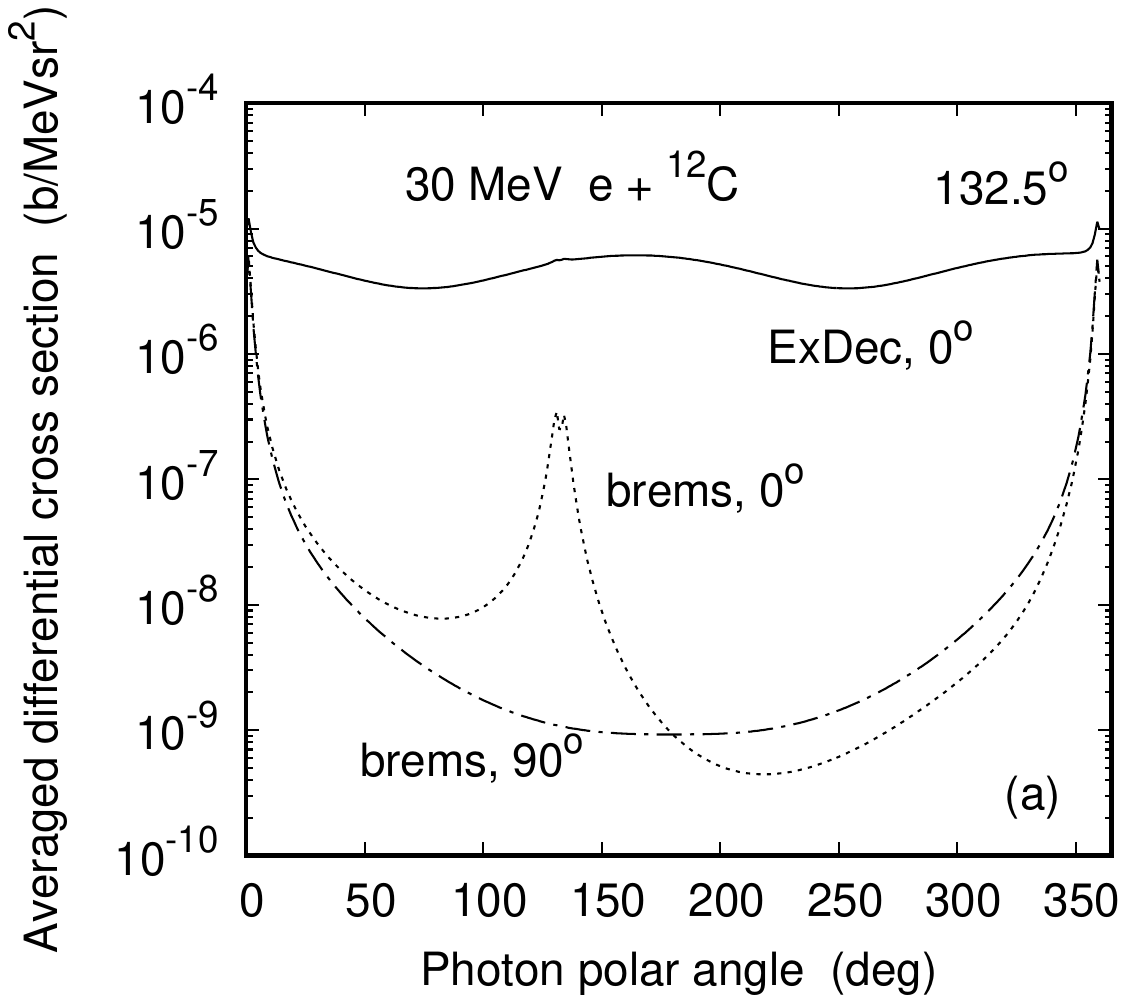}
\vspace{-1.5cm}
\vspace{-0.5cm}
\includegraphics[width=11cm]{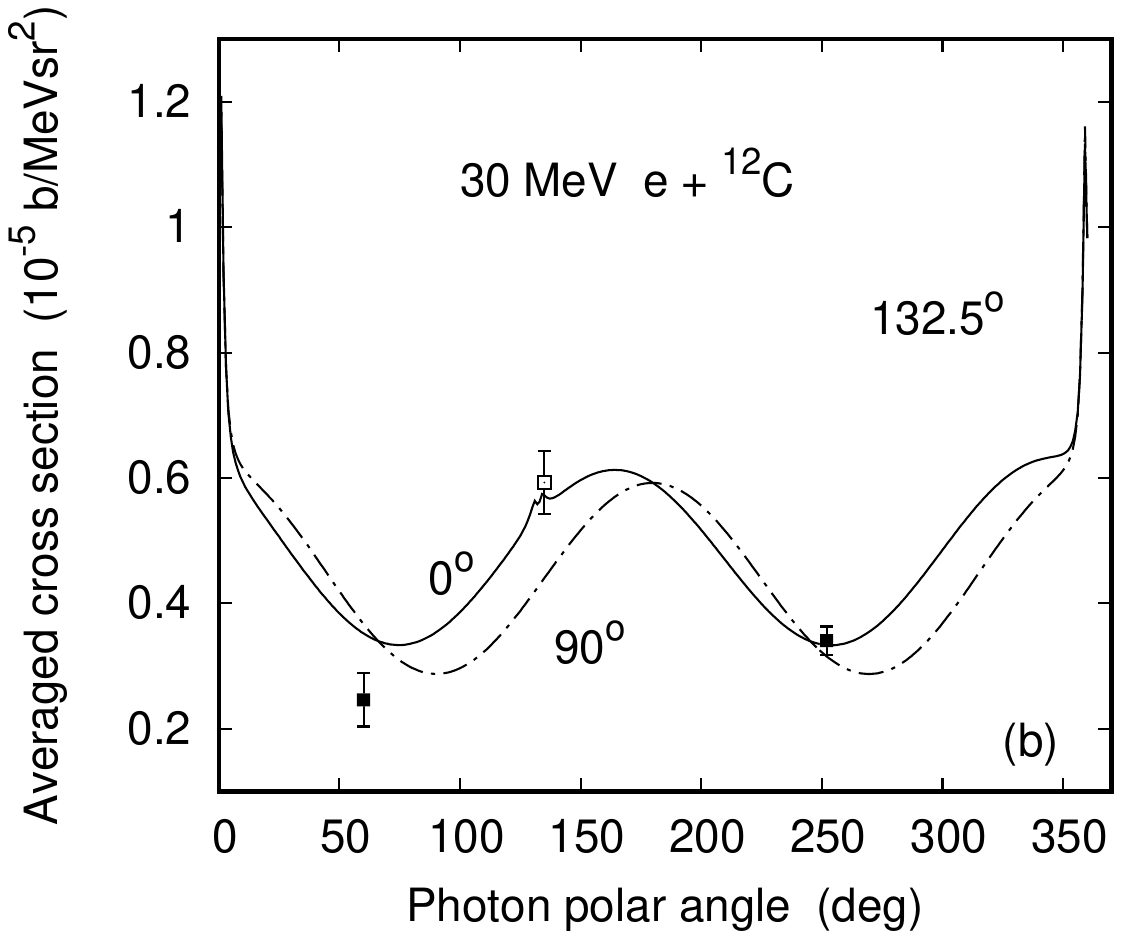}
\caption
{Triply differential cross section $\langle \frac{d^3\sigma}{d\omega d\Omega_k d\Omega_f} \rangle$ for the coincident excitation and decay 
(by photon emission of frequency $\omega$) of the $1_2^+$ state of $^{12}$C by 30 MeV electrons, averaged over the detector resolution $\Delta \omega/\omega=0.64\%$, as
a function of the photon angle $\theta_k$ at $\vartheta_f = 132.5^\circ$ (-----).
The QPM transition density is used, and bremsstrahlung is included.
In (a), the azimuthal photon angle is $\varphi_k=0.$ The bremsstrahlung cross section is shown separately for $\varphi_k=0 \;(\cdots\cdots)$ and $\varphi_k=90^\circ \;(-\cdot -\cdot -).$
In (b) comparison is made with the experimental data of Steinhilber \cite{St22}, normalized to theory at $252^\circ$.
Shown are the results for $\varphi_k=0$ (-----, $\blacksquare$) and for $\varphi_k=90^\circ\;(-\cdot - \cdot -, \;\square)$. 
}
\end{figure}

In order to apply the representation (\ref{2.9}) for the differential cross section and the corresponding sum rules, we have to get rid of the third variable in $M_{fi}^{tm}$.
This is done by deriving symmetry relations of these transition matrix elements, similar to those which hold for bremsstrahlung in coplanar geometry \cite{Jaku18}.
We  use the fact that $Y_{l_fm_l}(\hat{\bfk}_f)$ is real since $\bfk_f$, lying in the $(x,z)$-plane, has azimuthal angle zero,
such that $Y_{l_f-m_l}(\hat{\bfk}_f)=(-1)^{m_l}\,Y_{l_fm_l}(\hat{\bfk}_f).$
Further, upon reverting the sign of all magnetic quantum numbers which are summed over in (\ref{5.5}) and (\ref{5.7})
and using the symmetry relation $(j_1-m_1\,j_2 -m_2|\,j -m)=(-1)^{j_1+j_2-j}(j_1m_1\,j_2m_2|\,jm)$ of the Clebsch-Gordan coefficients, we obtain
$$M_{fi}^{tm}(-m_i,-m_s,-M_f)=(-1)^{1+m_i-m_s-M_f}$$
\begin{equation}\label{5.8}
\times\;M_{fi}^{tm}(m_i,m_s,M_f),
\end{equation}
provided $M_i=0 $ and $J_i=0$.
If $M_i=0$ but $J_i\neq 0$ there occurs an additional phase factor with an $L$-dependence, $(-1)^{J_i+L-J_f}$.
However, for $J_i\neq 0$ several values of $L$ can contribute to $M_{fi}$, such that there is no common phase factor, prohibiting
a relation of  type (\ref{5.8}).

So if there are just the two final states with $\pm M_f$ contributing to the cross section, the eight matrix elements $M_{fi}$ reduce to four independent ones.
In that case  we abbreviate $M_{fi}(m_i,m_s,M_f)$ according to (\ref{2.5}) and use (\ref{5.8}) for obtaining $M_{fi}(m_i,m_s,-M_f)$. Explicitly,
$$M_{fi}(\frac12,\frac12,-M_f)=(-1)^{1-M_f}M_{fi}(-\frac12,-\frac12,M_f)$$
$$=\;-H\,(-1)^{M_f}$$
$$M_{fi}(\frac12,-\frac12,-M_f)=G\,(-1)^{M_f},$$
$$ M_{fi}(-\frac12,\frac12,-M_f)=K\,(-1)^{M_f}$$
\begin{equation}\label{5.9}
M_{fi}(-\frac12,-\frac12,-M_f)=-J\,(-1)^{M_f}.
\end{equation}
The differential cross section is given by
\begin{equation}\label{5.10}
\frac{d\sigma}{d\Omega_f}(\bfzeta_i,\bfzeta_f)\,=\,N_0\sum_{M_f}\left| \sum_{m_im_s} a_{m_i}b^\ast_{m_s}\,M_{fi}(m_i,m_s,M_f)\right|^2.
\end{equation}
For the two final $M_f$ states,
it consists of two summands of the shape (\ref{2.6}), where in the second summand, $J$ is replaced by $-H$, $K$ by $G$, $G$ by $K$ and $H$ by $-J$ according to (\ref{5.9}),
since the common factor $(-1)^{M_f}$ plays no role in (\ref{5.10}).
Inserting (\ref{2.7}) and (\ref{2.8}) for $a_{m_i}$ and $b_{m_s}$ and using the definition (\ref{2.9}) for the polarization correlations, it turns out (because of
mutual cancellations) that only the seven parameters, $c_{33},\;c_{11},\;c_{22},\;c_{13},\;c_{20},\;c_{31}$ and $c_{02}$ are different from zero.
Their representation in terms of $J,\;G,\;K$ and $H$ is identical to the one from (\ref{2.10}), with $D_0$  given by (\ref{2.11}).
Due to the summation over the two $M_f$ states, the unpolarized cross section in (\ref{2.9}) is now given by $(\frac{d\sigma}{d\Omega_f})_0=N_0D_0.$
Thus the sum rule (\ref{3.1}), connecting just these seven nonvanishing parameters, holds also for the nuclear magnetic excitation.
Since, however, the remaining $c_{jk}$ are zero and not of the form given in (\ref{2.10}), the
 additional sum rules from (\ref{3.2}) are not valid.

For excitations with natural parity, $\pi_{if}=(-1)^L$, both Coulombic transitions (induced by the charge density $\varrho_L$) and transverse electric transitions
(mediated by the current transition densities $J_{L,L\pm 1}$) occur \cite{HB}.
By using the formalism from \cite{Jaku15} and reverting signs as demonstrated above, the resulting phase factor for the Coulombic transition  is $(-1)^{m_l+J_i+L-J_f}$ (with $m_l=m_i-m_s-M_f$), and for the transverse electric transition it is $(-1)^{m_l+J_i+\lambda-J_f+1}$ with $\lambda=L\pm 1$, coinciding with the Coulombic one.
Also here, $J_i=0$ is required, so that only one $L$ contributes to $M_{fi}$.
The symmetry relation for the transition amplitude $M_{fi}$, valid for both Coulombic and transverse electric excitations of a spin-zero nucleus, is given by
$$M_{fi}^{c,te}(-m_i,-m_s,-M_f)\,=\,(-1)^{m_i-m_s-M_f}$$
\begin{equation}\label{5.11}
\times \;M_{fi}^{c,te}(m_i,m_s,M_f).
\end{equation}
Since the difference to (\ref{5.8}) for the transverse magnetic transitions is just a global phase which vanishes upon insertion into the cross-section formula (\ref{5.10}), the sum rule (\ref{3.1}) is also valid in this case, if just the two states with $\pm M_f$ contribute.

\begin{figure}
\vspace{-1.5cm}
\includegraphics[width=11cm]{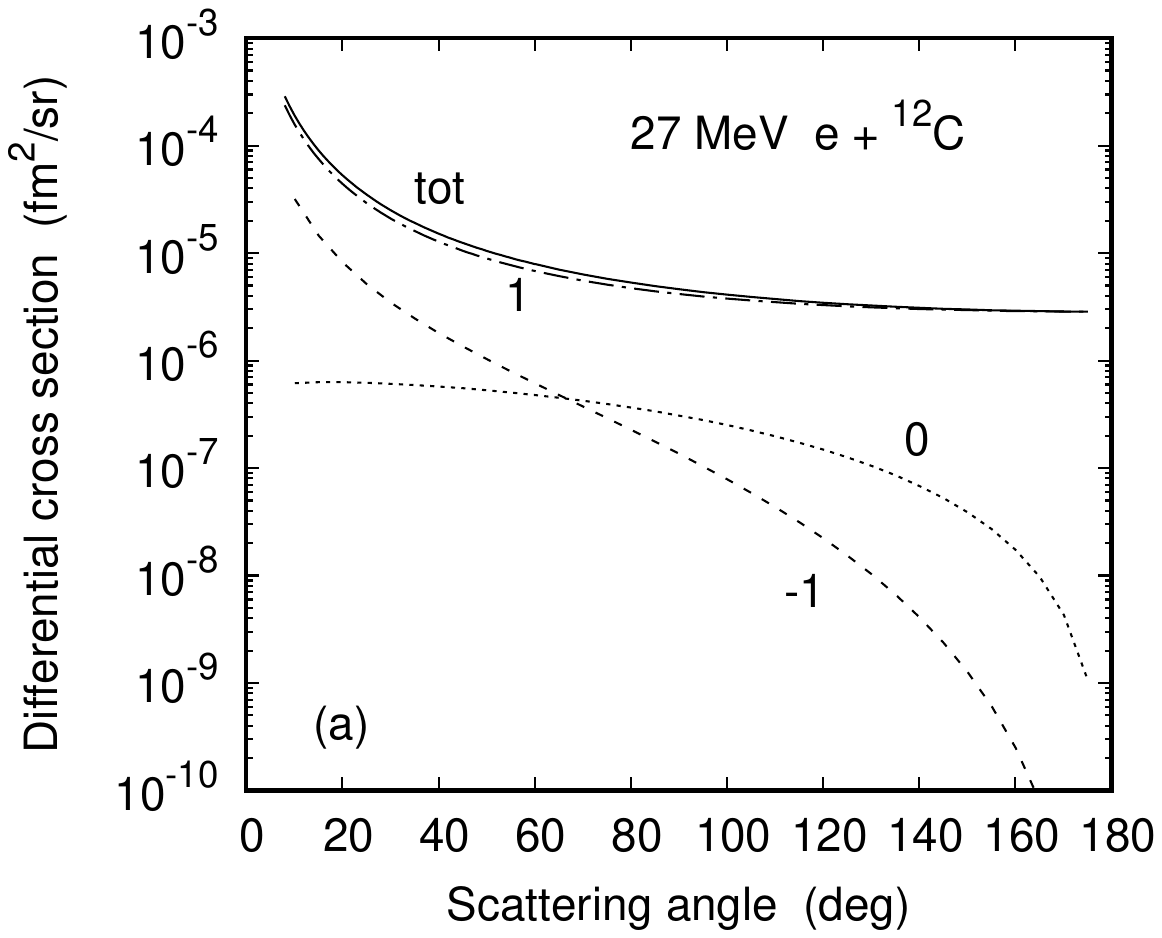}
\vspace{-1.5cm}
\vspace{-0.5cm}
\includegraphics[width=11cm]{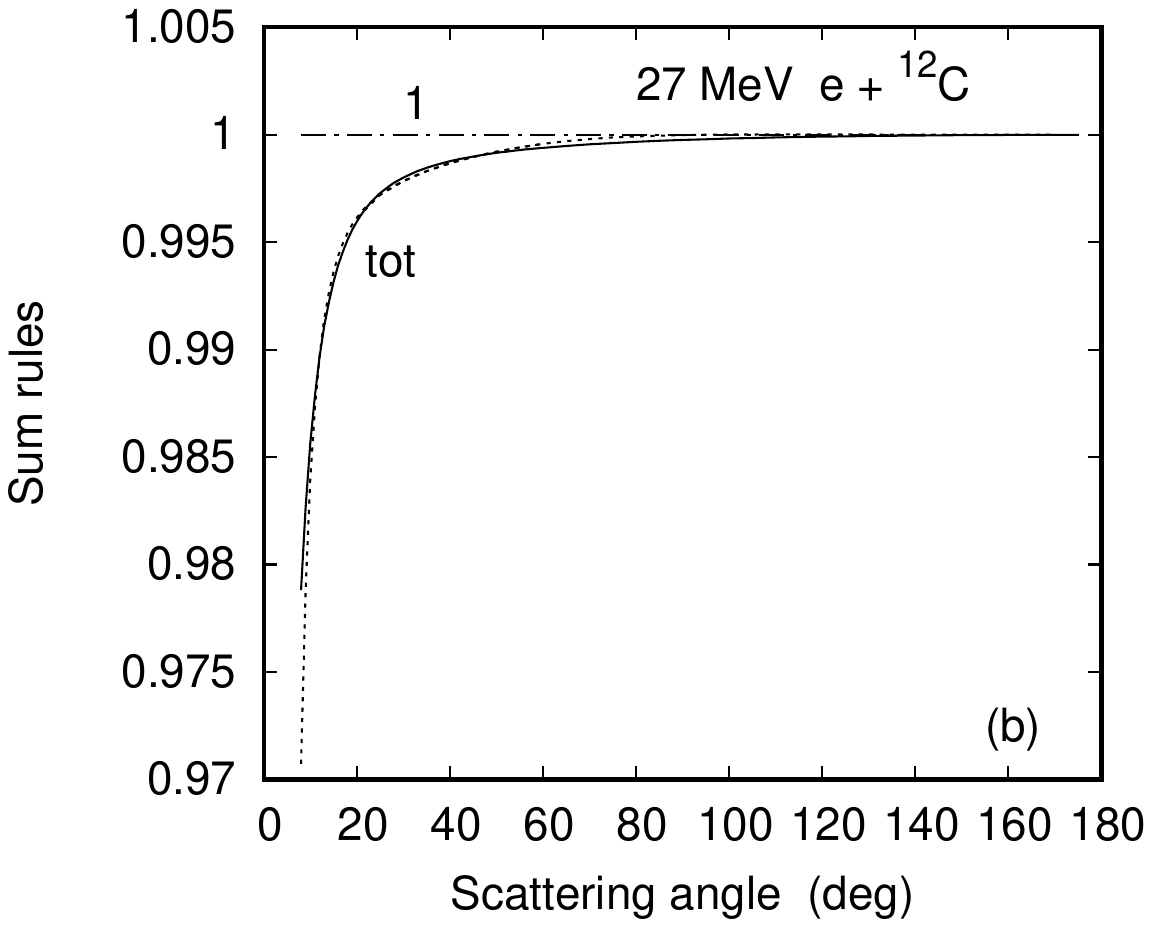}
\caption
{
(a) Differential cross section $(d\sigma/d\Omega_f)_0$ (-------) and (b)  sum rules  for the excitation of the 15.11 MeV state in 27 MeV $e+^{12}$C collisions as a function of scattering angle $\vartheta_f$. 
In (a) the partial cross sections relating to a fixed final state with $M_f=1 \;(-\cdot -\cdot -)$,  $M_f=0\;(\cdots\cdots)$ and $M_f=-1\; (----)$ are shown in addition.
In (b), $\Sigma_3$ for the excitation cross section (------, summed over $M_f$) and for the final state with $M_f=1\;(-\cdot -\cdot -)$ is displayed. 
$\Sigma_7 \;(\cdots\cdots,$ summed over $M_f$) is also shown.
Unity indicates that the sum rules are  valid.
}
\end{figure}

If instead,  the  final state $M_f=0$ strongly dominates the differential cross section, the symmetry relation (\ref{5.11}) turns into
\begin{equation}\label{5.12}
M_{fi}^{c,te}(-m_i,-m_s)\,=\,(-1)^{m_i-m_s}\,M_{fi}^{c,te}(m_i,m_s),
\end{equation}
from which one derives $H=J$ and $K=-G$, such that
$$c_{20}\,=\,-\;\frac{2\mbox{ Im }(J^\ast G)}{|J|^2\,+\,|G|^2},\qquad c_{33}\,=\,\frac{|J|^2\,-\,|G|^2}{|J|^2\,+\,|G|^2},$$
\begin{equation}\label{5.13}
c_{31}\,=\,-\;\frac{2 \mbox{ Re }(JG^\ast)}{|J|^2\,+\,|G|^2}.
\end{equation}
This leads to the three-term sum rule (\ref{4.5}).
As a consequence, the results for potential scattering (i.e. the validity of the seven-term sum rule (\ref{3.1}), $c_{20}=c_{02},\;c_{13}=-c_{31},\;c_{11}=c_{33}$ and $c_{22}=1)$ hold also in the $M_f=0$ case.
This is not true for a single final state with $M_f \neq 0$, although the three-term sum rule (\ref{1.2})  has been shown \cite{Jaku15} to be valid for general $M_i$ and $M_f$ 
 (provided only one initial and one final nuclear state contribute).
The respective proof was carried out by deriving the polarization correlations $S,\,L$ and $R$ explicitly from the cross-section differences (\ref{1.1}) 
by selecting a helicity (+) state ($b_\frac12 =\cos (\vartheta_f/2),\;b_{-\frac12}=\sin (\vartheta_f/2)$) for the scattered electron.
For such a final state, $R,\;S$ and $L$ are obtained by choosing $\bfzeta_i=\bfe_x,\;\bfe_y$ and $\bfe_z$, respectively \cite{Jaku12a}.

Due to the sign difference in (\ref{5.8}) for magnetic scattering, the relations in the case of an $M_f=0$ final state are different.
In place of (\ref{5.13}) one has
\begin{equation}\label{3.21}
c_{31}\;=\;\frac{2\mbox{ Re }(JG^\ast)}{|J|^2\,+\,|G|^2},
\end{equation}
while $c_{20}$ and $c_{33}$ are unchanged. Moreover, one gets $c_{20}=-c_{02},\;c_{13}=c_{31}, \;c_{11}=-c_{33}$ and $c_{22}=-1$.
The relations between $R,L$ and $c_{33},c_{11}$ become 
$$R\;=\;c_{33}\,\sin \vartheta_f\,-\,c_{31}\,\cos \vartheta_f$$
\begin{equation}\label{3.22}
L\;=\;c_{33}\,\cos \vartheta_f\,+c_{31}\,\sin \vartheta_f,
\end{equation}
while $S=c_{20}$ as for potential scattering.

\begin{figure}
\vspace{-1.5cm}
\includegraphics[width=11cm]{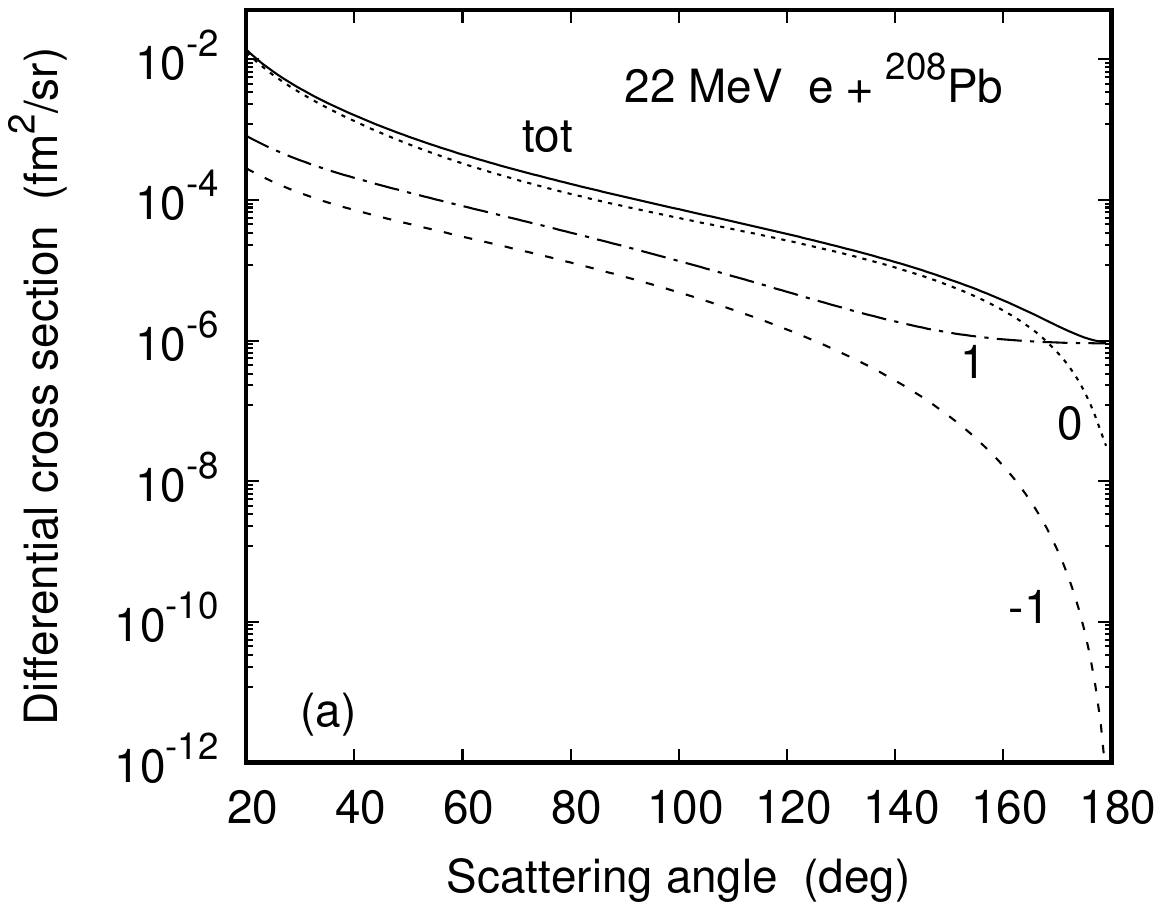}
\vspace{-1.5cm}
\vspace{-0.5cm}
\includegraphics[width=11cm]{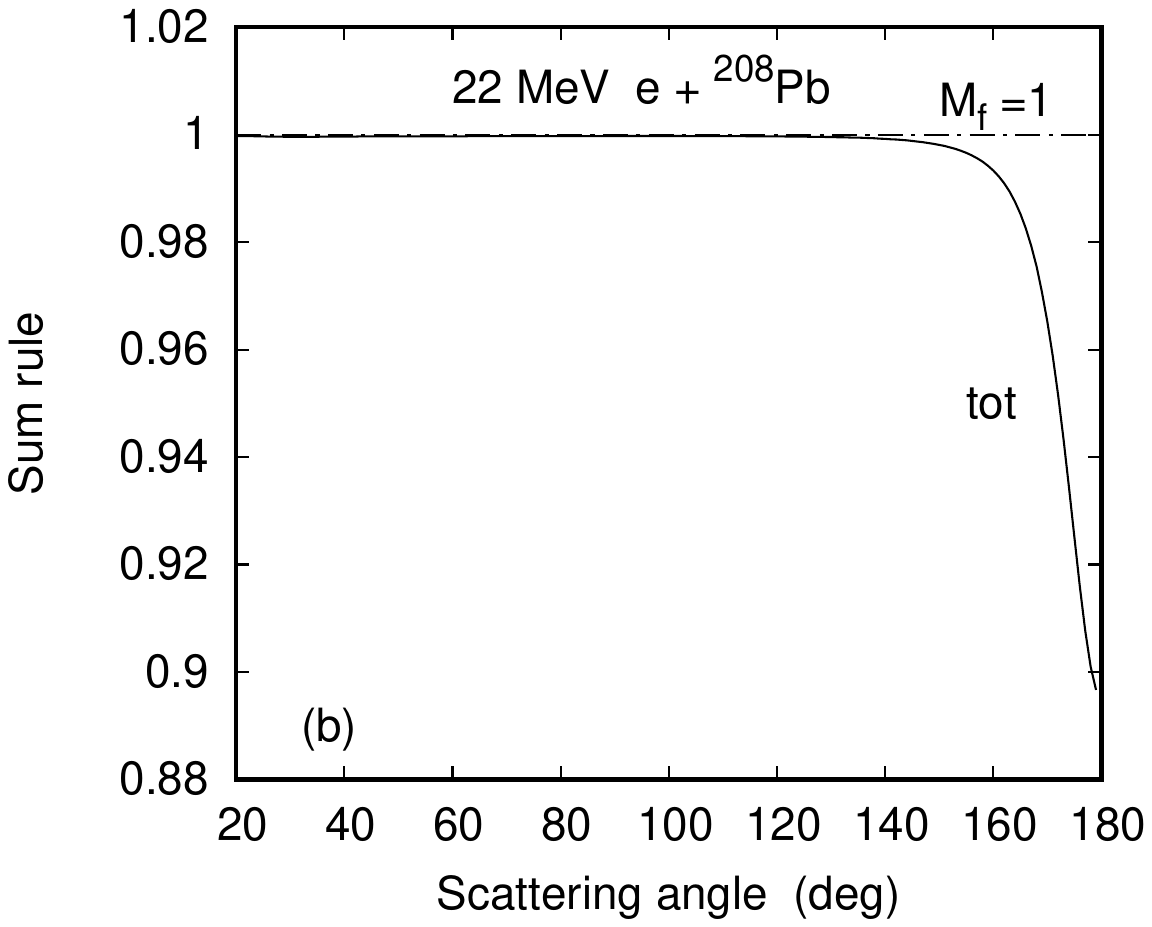}
\vspace{-1.5cm}
\vspace{-0.5cm}
\includegraphics[width=11cm]{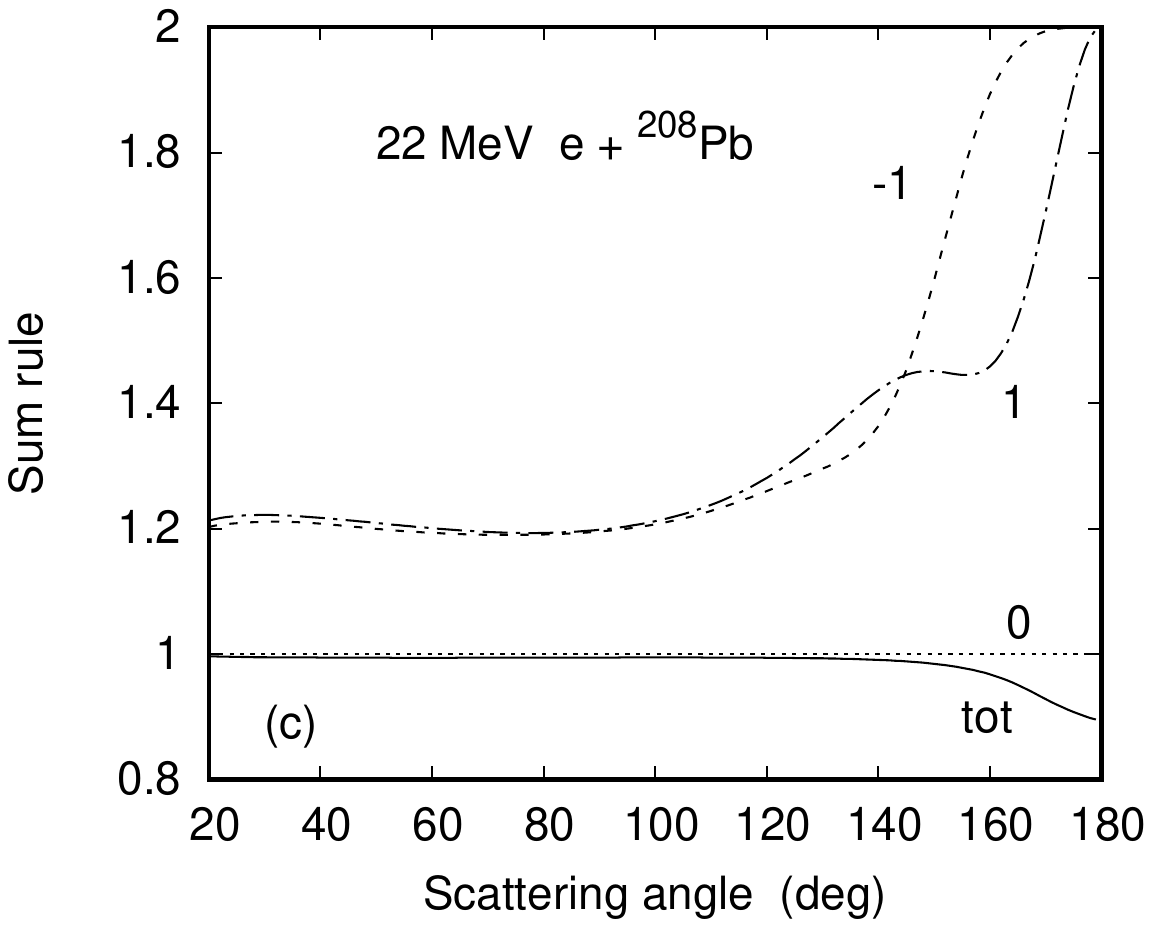}
\caption
{
Differential cross section and sum rules  for 22 MeV polarized electrons exciting the $1^-$ state at 5.512 MeV of $^{208}$Pb as a function of scattering angle $\vartheta_f$.
(a) $(d\sigma/d\Omega_f)_0$ from summing over all $M_f$ (-------) and its contributions from the $M_f=1 \;(-\cdot -\cdot -),\;M_f=0\;(\cdots\cdots)$ and $M_f=-1 \;(-----)$ state.
(b) Three-term sum rule $\Sigma_3$ from summing over all $M_f$ states (----------) and  from the $M_f=1$ contribution $(-\cdot - \cdot -)$.
(c) Seven-term sum rule $\Sigma_7$ for the contributions from the $M_f=1\;(-\cdot -\cdot -),\;M_f=0\;(\cdots\cdots)$ and $M_f=-1\;(-----)$ states, and from the sum over all $M_f$ (---------).
}
\end{figure}

\section{Specific examples for nuclear excitation}
\setcounter{equation}{0}

As examples, the magnetic excitation of the $1_2^+$ state of $^{12}$C at 15.11 MeV, and the electric  excitations of the $2_1^+$ state of $^{12}$C at 4.439 MeV
 and of the $1^-$ state of $^{208}$Pb at 5.512 MeV are considered.
For carbon, the differential cross section is tested against experiment.
Subsequently, the validity of the sum rules  is investigated.
Since helicity is conserved for an ultrarelativistic electron (when the electron's rest mass can be neglected), the respective parameter $L$ is unity up to $\vartheta_f \approx 180^\circ$ and hence $\Sigma_3 =1$.
In order to allow for a visible violation of the sum rules, a low impact energy has to be  chosen.

\subsection{Dipole excitation of $^{12}$C at 15.11 MeV}

The current transition density $J_{11}$, entering into (\ref{5.2}) for exciting the odd-parity $1_2^+$ state,  can be calculated within the random-phase approximation of the quasiparticle phonon model (QPM) \cite{So92,IP12,JP16}
and has been provided by Ponomarev (chosen to fit the experimental $B(M1\uparrow)$ strength distribution by using an effective $g$-factor, $g_s=0.9\,g_{s,free}$ \cite{Po19}).
Alternatively, $J_{11}$ has been  obtained from a Fourier-Bessel fit to experimental cross section data \cite{Deu83}. 
Fig.1 compares the spatial dependence of the transition density for the two models. They differ considerably for small $r_N$ which is important for large momentum transfers.

The differential cross section for the electron impact excitation of the $1_2^+$ state is calculated within the DWBA formalism given in section 3.3.
The nuclear potential of $^{12}$C is generated from the Fourier-Bessel expansion of the ground-state charge distribution \cite{deV}, and the
scattering eigenstates are obtained by solving the Dirac equation with the help of the Fortran code RADIAL by Salvat et al. \cite{Sal}.
Fig.2a shows the angular distribution  in comparison with the experimental data of Deutschmann et al. \cite{Deu83}.
By definition, the Fourier-Bessel density describes experiment very well, while the QPM model slightly overestimates the cross section at the larger scattering angles.

The energy distribution of the excitation cross section at a scattering angle of $135^\circ$ is displayed in Fig.2b. The experimental data by Proca and Isabelle \cite{PI68} are reasonably well described by theory.
At large momentum transfers (collision energies well above 100 MeV), the QPM model for $J_{11}$ is superior to the experimental fit.

There exist also new (relative) experimental data on the coincident nuclear excitation and decay (ExDec) process for the $1_2^+$ state, measured at the S-DALINAC accelerator \cite{St22}.
The theoretical formalism for obtaining the triply differential cross section for this process is again based on the DWBA and is described in \cite{HR66,JP17}.
A competitive process is the emission of bremsstrahlung by the beam electron at a photon frequency $\omega$ matching the nuclear excitation energy, 
which has to be added coherently.

We recall that in the calculations the $z$-axis is taken along the beam direction and not along the momentum transfer (as done in early papers or in \cite{St22}).
Thus the azimuthal photon angle $\varphi_k$ is the angle between the $x$-axis and the projection of the photon momentum $\bfk$ onto the ($x,y)$ plane.

\begin{figure}
\vspace{-1.5cm}
\includegraphics[width=11cm]{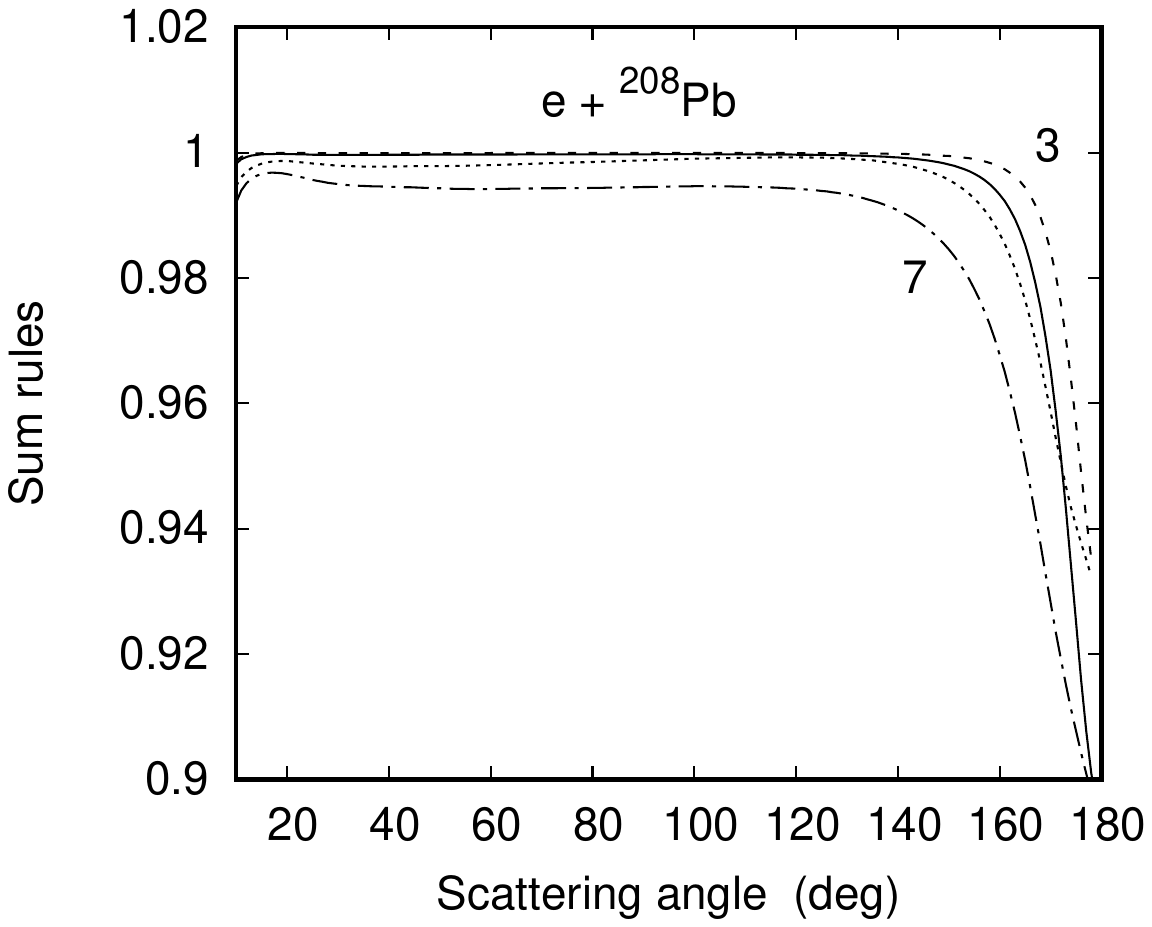}
\caption
{
Sum rules $\Sigma_3$ and $\Sigma_7$ for 22 MeV and 30 MeV electrons exciting the $1^-$ state of $^{208}$Pb (summed over $M_f$) as a function of scattering angle $\vartheta_f$.
---------, $\Sigma_3$ for 22 MeV; $-----,\; \Sigma_3$ for 30 MeV; $-\cdot -\cdot -,\;\Sigma_7$ for 22 MeV; $\cdots\cdots,\;\Sigma_7$ for 30 MeV.
}
\end{figure}

Due to the tiny decay width $\Gamma = 35.9$ eV of the $1_2^+$ state to the ground state, the measured intensity of the decay photons depends crucially on the energy resolution of the photon detector.
Fig.3a shows the differential cross section  at a scattering angle of $132.5^\circ$ as a function of the polar photon angle $\theta_k$,
averaged over the resolution of the photon detectors which is about 0.64\% (corresponding to a peak FWHM of 97 keV). 
The dipole pattern corresponding to the decay of an $L=1$ state is clearly seen.
Separately shown is  the differential cross section for bremsstrahlung emission, which is peaked near $\theta_k=0$ and $360^\circ$, and for a small azimuthal angle $\varphi_k$ 
also at $\theta_k\approx \vartheta_f$.
Only in the  vicinity of these  angles is the ExDec intensity influenced by bremsstrahlung.
In Fig.3b comparison is made with experiment, measured at $\varphi_k=0$ and $\varphi_k=90^\circ$, and normalized to theory at $\theta_k=252^\circ$.

Let us now turn to the investigation of the sum rules for the $1_2^+$ state.
All parameters of the seven-term sum rule are calculated from the relative cross-section differences as obtained with the parametrization (\ref{2.9}) and a suitable choice of $\bfzeta_i$ and $\bfzeta_f$,
\begin{equation}\label{4.1}
c_{jk}\,=\,\frac{\sum_{M_f}|W_{fi}(\bfzeta_i,\bfzeta_f)|^2\,- \sum_{M_f} |W_{fi}(-\bfzeta_i,\bfzeta_f)|^2}{\sum_{M_f}|W_{fi}(\bfzeta_i,\bfzeta_f)|^2\,+\sum_{M_f}|W_{fi}|(-\bfzeta_i,\bfzeta_f)|^2},
\end{equation}
where for $c_{20}$ and $c_{02}$ an additional sum over $\bfzeta_f$, respectively average over $\bfzeta_i$, has to be carried out.
In contrast, the parameters $S$ and $L$ are calculated from the explicit formulae given in \cite{Jaku15}, which are equivalent to related expressions in (\ref{2.10}).

Fig.4a displays the angular distribution of the excitation cross section at 27 MeV collision energy, together with the subshell contributions pertaining to the $J_f=1$ final state with  fixed $M_f \in \{0,\pm 1\}$.
While at small angles both $M_f=\pm 1$ states contribute notably to the excitation cross section, it is predominantly $M_f=1$ in the backward hemisphere.
Correspondingly, the three-term sum rule (\ref{1.2}) is well satisfied at large angles, but is violated in the forward direction.
A similar behaviour is found for the seven-term sum rule (Fig.4b).
Note that, if only one single $M_f$ is allowed throughout, it is verified in this figure that $\Sigma_3$ is  unity.

\subsection{Dipole excitation of $^{208}$Pb at 5.512 MeV}

We are now considering the excitation of an even-parity dipole state, measured in $^{208}$Pb$(\gamma,\gamma')$ reactions \cite{Pi09}.
The corresponding transition densities $\varrho_1,\;J_{12}$ and $J_{10}$ are obtained from the QPM model and are displayed in \cite{JP16}.

Fig.5a shows the angular  distribution of the excitation cross section at 22 MeV, together with the corresponding subshell contributions.
In contrast to the odd-parity excitation, the $M_f=0$ state largely dominates the cross section.
Correspondingly, both the three-term and the seven-term sum rules are approximately valid for $\vartheta_f \lesssim 140^\circ$ (Fig.5b,c), beyond which the $M_f=1$ state comes into play and finally takes over.
While it is demonstrated (for the case $M_f=1$) that the three-term sum rule holds at all angles if just one specific final $M_f$ state is allowed, the seven-term sum rule is merely valid for $M_f=0$ 
 (Fig.5c).

When the collision energy is increased, both sum rules deviate less from unity in the whole angular regime if it is summed over all $M_f$ states.
This is shown in Fig.6 for the impact energies 22 MeV and 30 MeV.
In contrast to the results for the $^{12}$C target,  there is a considerable difference between $\Sigma_3$ and $\Sigma_7$ for $^{208}$Pb.

\begin{figure}
\vspace{-1.5cm}
\includegraphics[width=11cm]{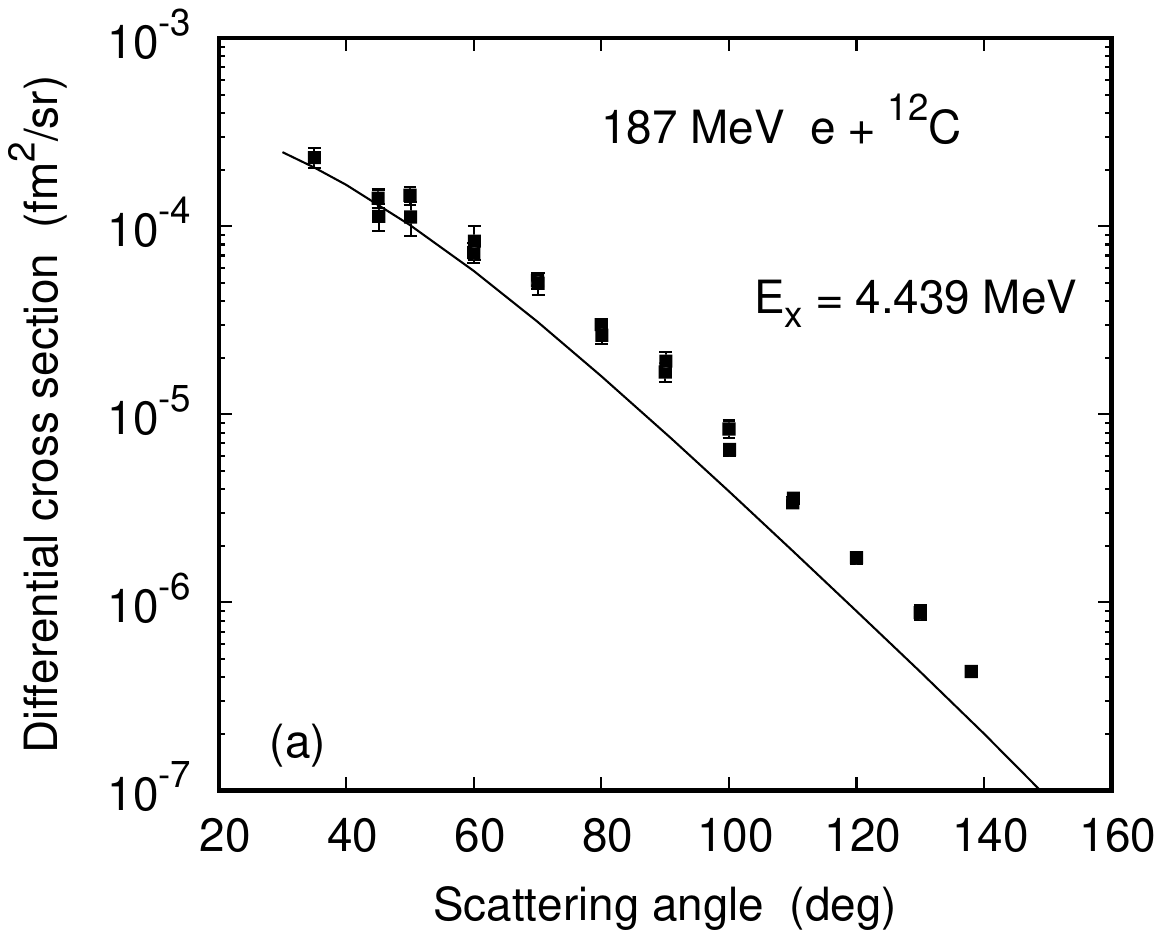}
\vspace{-1.5cm}
\vspace{-0.5cm}
\includegraphics[width=11cm]{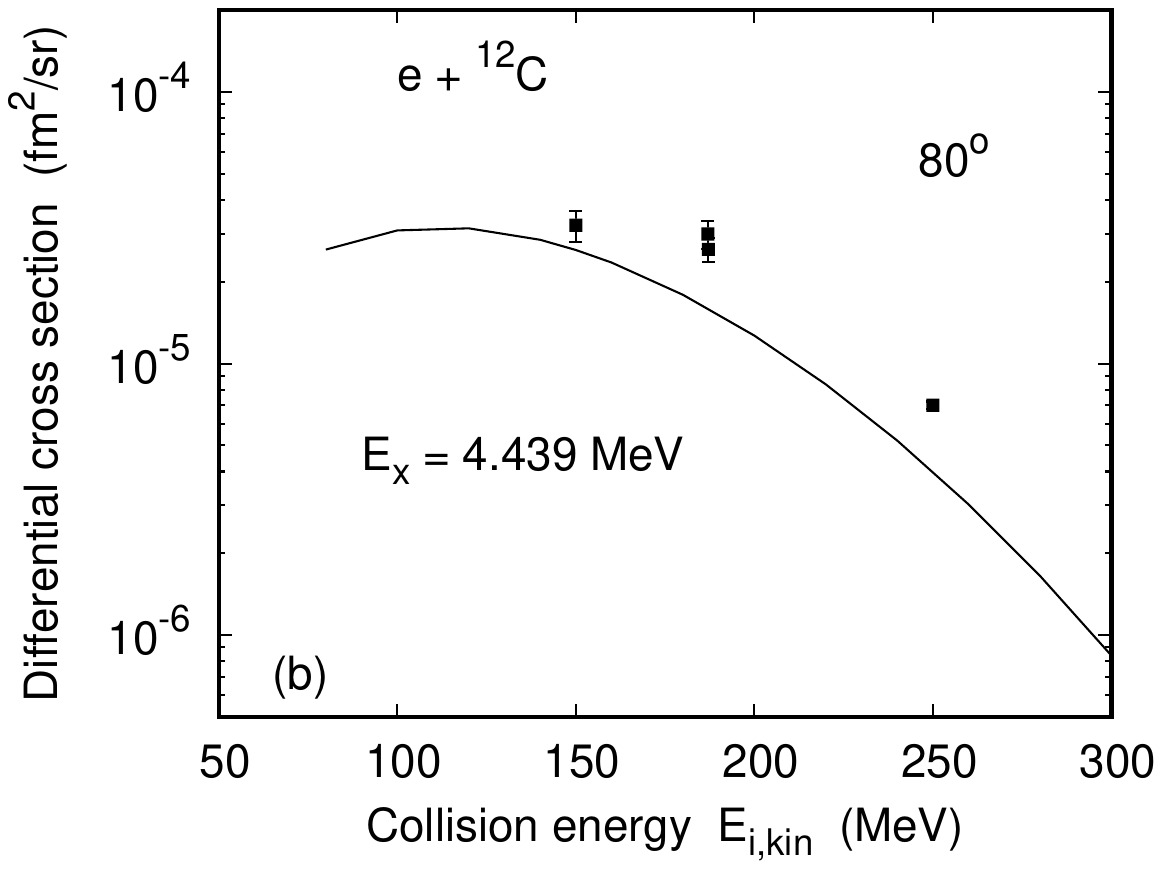}
\caption
{
Differential cross section $d\sigma/d\Omega_f$ (-------) for the excitation of the 4.439 MeV state in $e+^{12}$C collisions (a) at 187 MeV collision energy as a function of scattering angle $\vartheta_f$, and (b) at $\vartheta_f=80^\circ$ as a function of collision energy.
In (a), comparison  is made with the experimental data of Fregeau ($\blacksquare$, \cite{Fr56}) and in (b) with the data ($\blacksquare$) of Fregeau at 150 and 187 MeV \cite{Fr56} and of Crannell and Griffy at 250 MeV \cite{CG64}.
}
\end{figure}

\subsection{Quadrupole excitation of $^{12}$C  at 4.439 MeV}

The transition densities $\varrho_2,\;J_{23}$ and $J_{21}$ for the even-parity $2_1^+$ excitation are again obtained from the QPM model \cite{Po19}, and their spatial dependence is provided in \cite{JP20}.
In comparison with experiment \cite{Fr56,CG64}, the angular distribution of the differential excitation cross section at a collision energy of 187 MeV is shown in Fig.7a,
and the energy distribution at a scattering angle of $80^\circ$ is depicted in Fig.7b.
Theory underestimates the data by  mostly a factor of two, but the global dependence on energy and angle is reasonably well described.

Fig.8a diplays the excitation cross section of the $2_1^+$ state at 30 MeV electron impact, as well as the separate contributions from the states with fixed $M_f$.
It is seen that the states with the same modulus of $M_f$ are excited with nearly equal probability in the forward hemisphere,
while, like for the  excitation of  the dipole states, the $M_f=1$ state is dominant at the backmost angles.
The $M_f=0$ state has a low excitation probablility in the forward hemisphere, with a deep minimum at $64^\circ$, but it is dominant between $120^\circ$ and $170^\circ$.

The polarization correlations $S$ and $c_{20}$ for perpendicularly polarized beam electrons are shown in Fig.8b and coincide at all angles.
The parameter $c_{02}$ for perpendicularly polarized scattered electrons obeys $c_{02} =  c_{20}$ to an accuracy better than  1\% for $\vartheta_f \gtrsim 100^\circ$, except for some deviation at the smaller angles.
Note the deep backward minimum of $S$, which is characteristic for the Sherman function at small collision energies.

Fig.8c displays the
polarization correlation $L$.
It deviates from unity only above $100^\circ$, allowing for the spin flip of a helicity (+) electron just at the backmost angles.
Also shown is the parameter $c_{22}$ which notably differs from unity  above $80^\circ$. Additionally  included are the 
sum rules $\Sigma_3$ and $\Sigma_7$.
One has $\Sigma_3 \approx \Sigma_7$ and the sum rules hold up to $150^\circ$. For higher collision energies, the deviations of $\Sigma_3 $ and $\Sigma_7$ from unity become smaller.
Since there are five final $M_f$ states such that there is basically no angular region where just one or two of them (the ones with $\pm M_f$) are important, the validity or violation of the sum rules
cannot be linked to a respective behaviour of the cross section as in the case of the dipole excitations.

\begin{figure}
\vspace{-1.5cm}
\includegraphics[width=11cm]{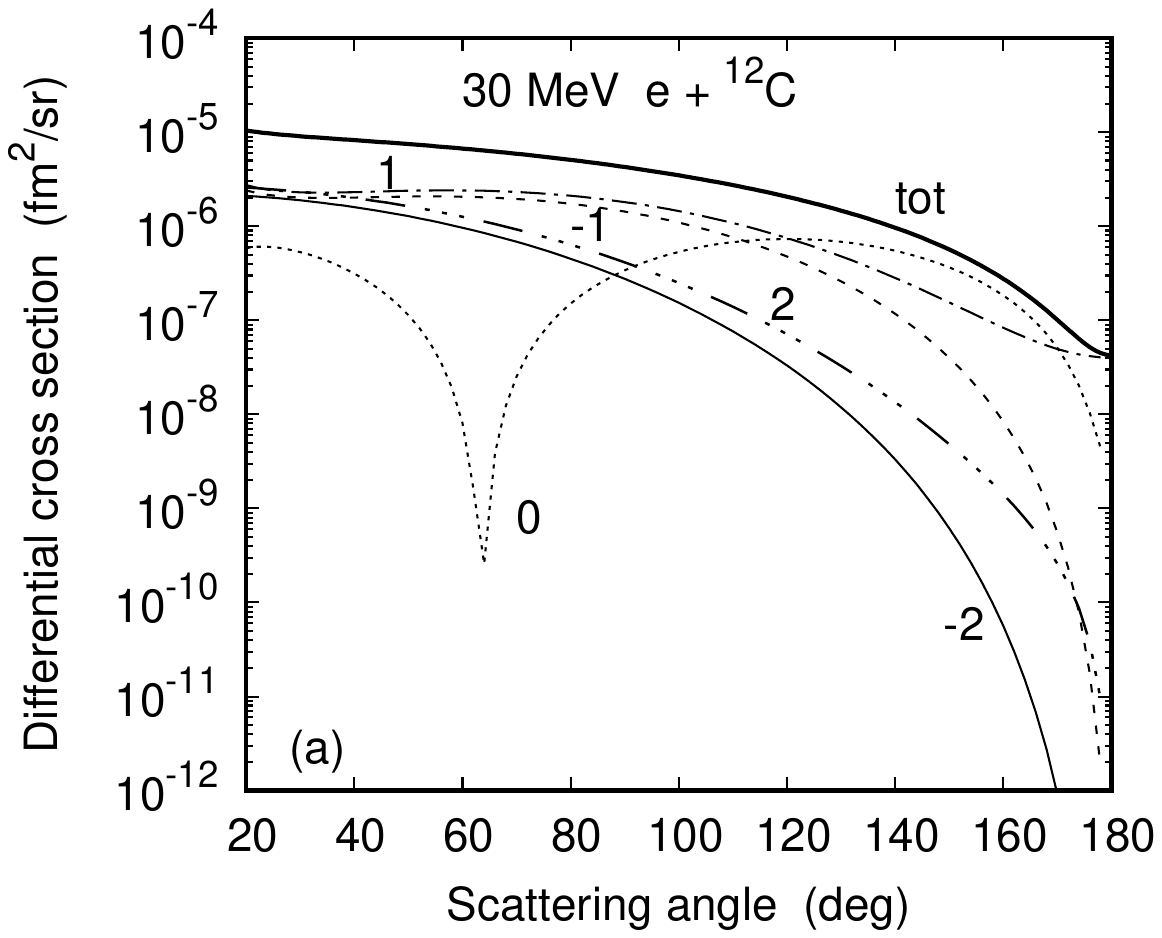}
\vspace{-1.5cm}
\vspace{-0.5cm}
\includegraphics[width=11cm]{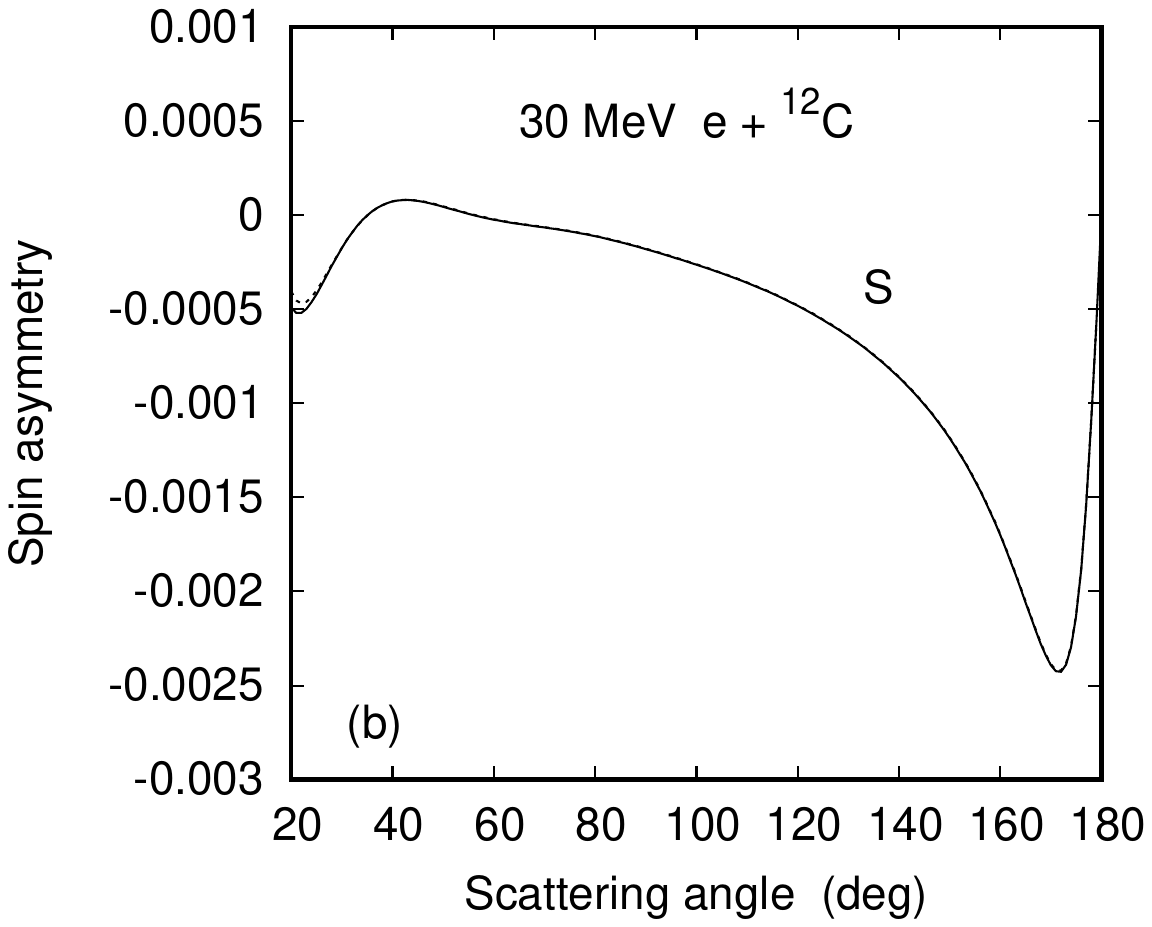}
\vspace{-1.5cm}
\vspace{-0.5cm}
\includegraphics[width=11cm]{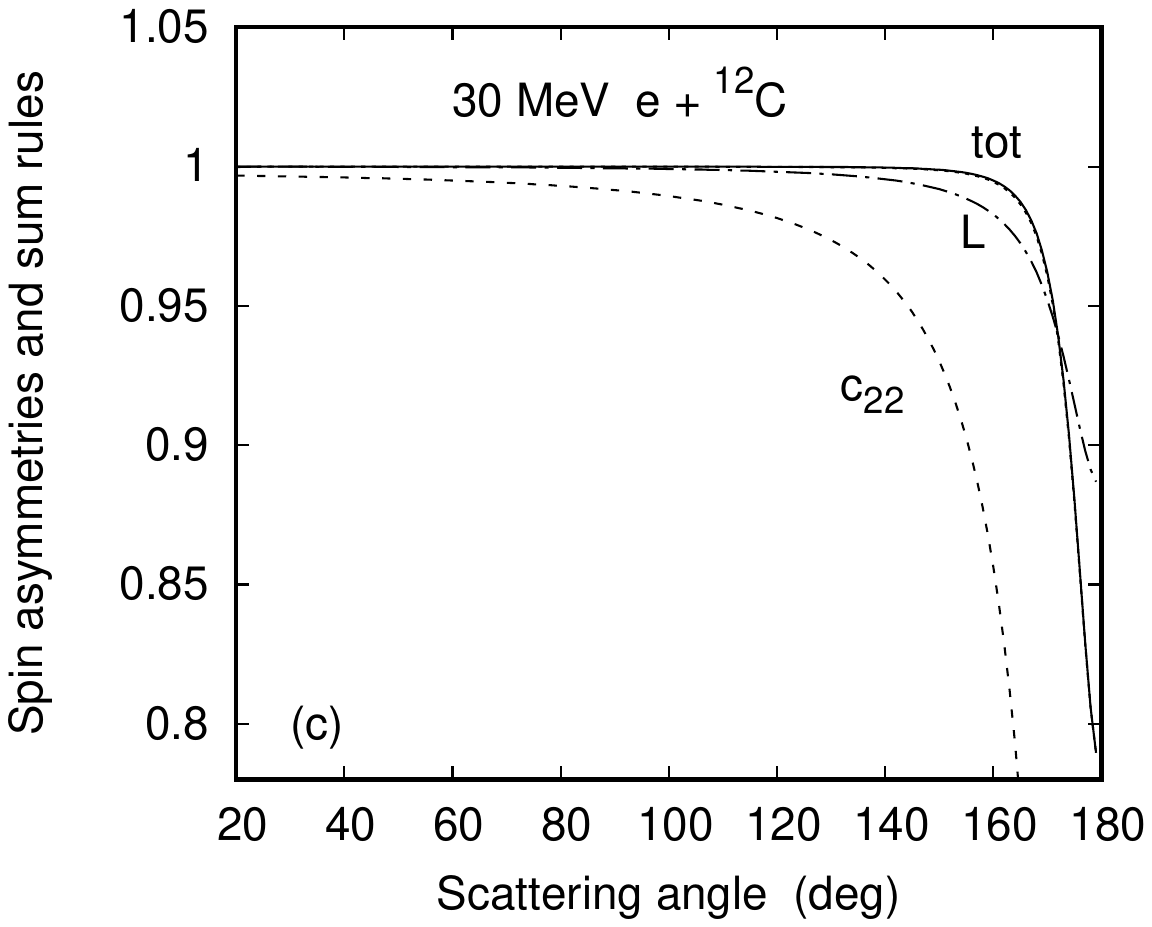}
\caption
{
Differential cross section, spin asymmetries and sum rules for 30 MeV polarized electrons exciting the $2_1^+$ state of $^{12}$C as a function of scattering angle $\vartheta_f$.
(a) $(d\sigma/d\Omega_f)_0$ from summing over all $M_f$ states (----------, thick line) and its contributions from the $M_f=1$ state ($-\cdot -\cdot -),$ the $M_f=-1$ state $(-----)$, the $M_f=0$ state $(\cdots\cdots$), the $M_f=2$ state ($-\cdots -\cdots-$) 
and the $M_f=-2$ state (--------, thin line).
(b) Polarization correlations $S$ (---------) and $ c_{02}\; (\cdots\cdots)$ from summing over all $M_f$. The difference between $S$ and $c_{20}$ is smaller than $10^{-8}\%$ and hence not visible in the plot.
(c) Polarization corelations $L \;(-\cdot -\cdot -)$ and $c_{22}\;(-----)$ as well as the sum rules $\Sigma_3$ (-------) and $\Sigma_7 \;(\cdots\cdots)$.
}
\end{figure}

\section{Numerical accuracy of the sum rules}
\setcounter{equation}{0}

From theory it follows that the three-term sum rule is strictly valid if only a single final state (with fixed $M_f$) is taken into account.
The numerical verification of $\Sigma_3=1$ in that case holds at an accuracy better than $10^{-12}$. This can be related to the fact that $\Sigma_3$ is calculated from the closed expressions (\ref{5.13}) of $c_{20},\;c_{33}$ and $c_{31}$,
where the accuracy of each of these three parameters (in terms of $J$ and $G$) is irrelevant.

We recall that the parameters of the seven-term sum rule are calculated from (\ref{4.1}).
Lacking a closed expression as for $c_{20},\;c_{33}$ and $c_{31}$ (in the case of a single $M_f$ state), there appear numerical inaccuracies when forming the difference of nearly equally large summands.
This was already shown to be the case for a seven-term sum rule in bremsstrahlung. There, inaccuracies up to 20\% had been encountered in its numerical verification for the heaviest nuclei \cite{Jaku16}.

For nuclear excitation
the  situation is considered
where only the two  $\pm M_f$ states are included in the cross section, since in that case theory predicts that the seven-term sum rule (\ref{3.1}) is
valid.
Fig.9a displays its numerical verification for 30 MeV electrons colliding with $^{12}$C and $^{208}$Pb, when the $M_f=\pm 1 $ states are summed.
For the carbon $2_1^+$ state, the numerical error is quite small ($\lesssim 10^{-4}$), while for the $1_2^+$ state it is up to 0.5\% and slightly more at $\vartheta_f < 20^\circ$.
For the $1^-$ state of $^{208}$Pb, on the other hand, the inaccuracy is much larger, up to 4\% at the smallest angles considered.
Included in this figure is the result for the three-term sum rule  in the case of the $2_1^+$ state of $^{12}$C.
This sum rule is violated above $140^\circ$.
 However, near $180^\circ$ where the $M_f=-1$ state is strongly suppressed  (such that only one final state contributes), 
the three-term sum rule holds again.

In order to compare the results for the $L=2$ state of $^{12}$C with a corresponding $L=2$ state of a heavier nucleus at the same collision energy,
the $2_2^+$ excitation of $^{92}$Zr at 1.847 MeV is considered (which is closer in energy and current transition densities than the $2_1^+$ state of $^{92}$Zr at 0.9345 MeV). The respective transition densities are provided in \cite{JP17}.

In Fig.9b the results for $\Sigma_7$ arising from the $M_f=\pm 1$ states in comparison with the $M_f=\pm 2$ states
for the quadrupole excitation of $^{12}$C and of  $^{92}$Zr are shown.
It is obvious that the numerical error increases both with target charge and with $|M_f|$, being particularly large in the backward hemisphere.
At small angles only the $|M_f|=2$ states lead to a notable deviation from unity.

\begin{figure}
\vspace{-1.5cm}
\includegraphics[width=11cm]{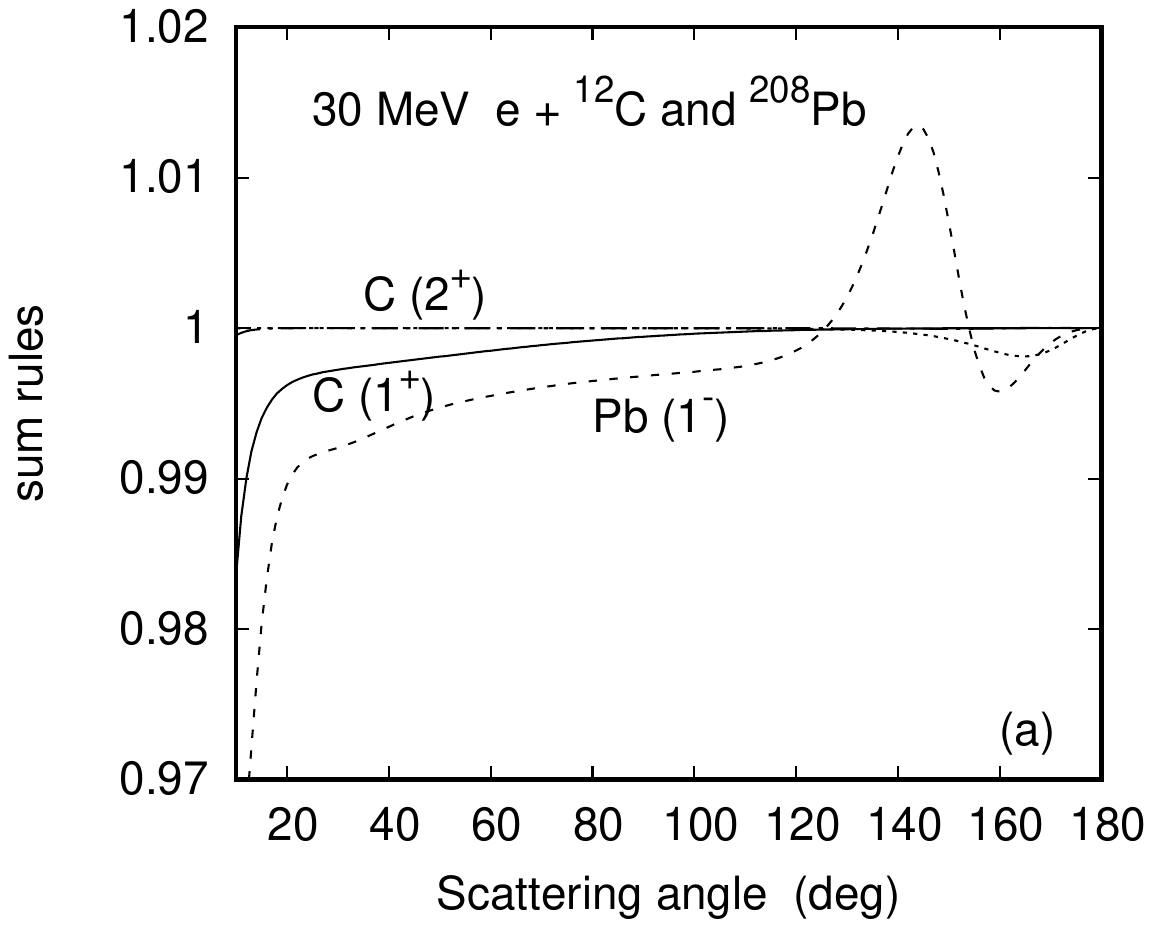}
\vspace{-1.5cm}
\vspace{-0.5cm}
\includegraphics[width=11cm]{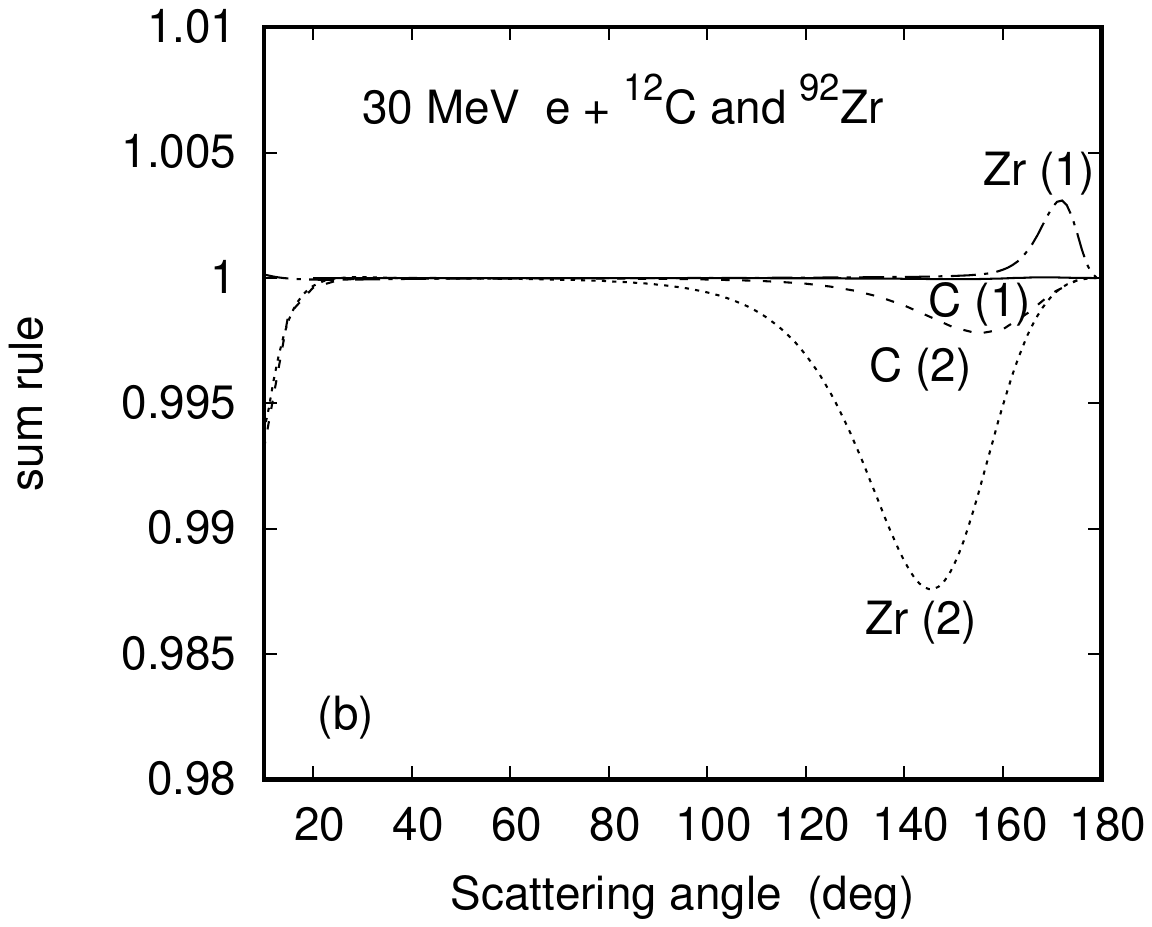}
\caption
{
Sum rule $\Sigma_7$ for 30 MeV electron impact as a function of scattering angle $\vartheta_f$ if only the two $\pm M_f$ states are included.
(a) for $|M_f|=1  $ and $e+^{12}$C, $E_x=4.439$ MeV $(-\cdot -\cdot -)$,  $e+^{12}$C, $E_x=15.11$ MeV (---------) and  $e+^{208}$Pb, $E_x=5.512$ MeV $(-----)$. The sum rule $\Sigma_3$ for the $2_1^+$ state of $^{12}$C is also shown $(\cdots\cdots)$.
(b)  for the $2_1^+$ state of $^{12}$C (---------, $|M_f|=1; \; -----,\;|M_f|=2$)
and for the $2_2^+$ state of $^{92}$Zr ($-\cdot -\cdot -,\;|M_f|=1;\;\cdots\cdots,\;|M_f|=2$).
}
\end{figure}

\section{Conclusion}

By using the correspondence between the representation of the polarization correlations pertaining to polarized bremsstrahlung emission and those
pertaining to radiationless electron scattering, three seven-term sum rules were established for the polarization correlations relating to the latter process.
In the special case of potential scattering, the first of these sum rules (which corresponds to an unpolarized photon in the bremsstrahlung process) reduces to the well-known three-term sum rule, while the other two sum rules are trivially satisfied.

For the case of nuclear excitation, exemplified by dipole and quadrupole excitations of spin-zero nuclei like $^{12}$C and $^{208}$Pb, the DWBA scattering theory together with the QPM prescription for the nuclear structure
was tested against experimental differential cross sections, verifying the theoretical energy and angular distributions.
Low impact energies were chosen such that the violation of the three-term sum rule (necessitating $L\neq 1)$ is not confined to a too small angular region.

Symmetry relations between the transition matrix elements pertaining to a nuclear excited state with  magnetic quantum number $\pm M_f$
reduce the fifteen polarization correlations of the general theory to seven.
In the case where only one single $M_f$ state is selected, the three-term sum rule is always valid,
while the above-mentioned seven-term sum rule holds only for $M_f=0$, and there with the identical interrelations of the polarization correlations as for potential scattering.
The violation of the seven-term sum rule for $M_f \neq 0$ is related to the coupling  of this specific nuclear state to the electronic spin states, 
such that the $M_f$ in the transition matrix element $M_{fi}(m_i,m_s,M_f)$ cannot be treated as an independent variable.

If, on the other hand, both final states with $\pm M_f$ are included in the differential cross section, the seven-term sum rule is valid.
However, the three-term sum rule is violated since there is no additional condition on the transition matrix elements which could cause any interrelation between the seven polarization correlations.

For the dipole excitations with angular regions where one final $M_f$ state is strongly dominating (while the contribution of the remaining two states is irrelevant) 
it could be shown that the three-term sum rule is violated as soon as some additional final state comes into play.
For the quadrupole excitation with as many as five final $M_f$ states an interrelation between the validity of this sum rule and the contribution of various final states
to the differential cross section could not be found.
Neither could any relation between the seven-term sum rule and the behaviour of the cross section be established for any of the targets.

For the  case where only the two $\pm M_f$ states are included in the cross section, the verification of the seven-term sum rule was tested numerically.
It was found that the deviations from $\Sigma_7=1$, being a measure of the inaccuracy of the numerical evaluation of the transition amplitudes, increase with $|M_f|$ and with the charge number of the target nucleus, but decrease with collision energy except possibly at the backmost angles.

When the dipole excitation of a spin $\frac12^-$ nucleus (such as $^{207}$Pb) to a $\frac12^+$ state is considered, there are just the two final $M_f=\pm \frac12$ states which contribute to the cross section.
However, one has $M_i \neq 0$ such that $M_i$ enters as an additional parameter into the transition matrix element.
Even with the corresponding symmetry relations, eight different matrix elements remain. This leads to 63 nontrivial polarization correlations like in the case of noncoplanar bremsstrahlung emission.
For those, no sum rule has yet been established.
Hence an experimental verification of the above results seems only possible for dipole excitations of spin-zero nuclei at low collision energies.

\vspace{0.5cm}

\noindent{\small \bf ACKNOWLEDGMENT}

I would like to thank V.Yu.Ponomarev for calculating the nuclear transitions densities.


\vspace{1cm}

\end{document}